\documentclass[preprint]{aastex}

\newcommand{\lsim}{\stackrel{<}{{}_\sim}}

\newcommand\xion[2]{#1$\;${\small\rmfamily#2}\relax}% Added by arXiv admin to allow commands like \ion{Ca}{K} to process without user intervention

\begin{document}

\title{Spectral Classification of Quasars in the Sloan Digital Sky Survey: Eigenspectra;
 Redshift and Luminosity Effects}

\author{C.~W.~Yip\altaffilmark{1},
A.~J.~Connolly\altaffilmark{1}, 
D.~E.~Vanden~Berk\altaffilmark{1}, 
Z.~Ma\altaffilmark{2}, 
J.~A.~Frieman\altaffilmark{2}, 
M.~SubbaRao\altaffilmark{2,7}, 
A.~S.~Szalay\altaffilmark{3}, 
G.~T.~Richards\altaffilmark{4},
P.~B.~Hall\altaffilmark{4},
D.~P.~Schneider\altaffilmark{5}, 
A.~M.~Hopkins\altaffilmark{1}, 
J.~Trump\altaffilmark{5},
J.~Brinkmann\altaffilmark{6}} 

\altaffiltext{1}{Department of Physics and Astronomy, University of Pittsburgh, Pittsburgh, PA 15260} 

\altaffiltext{2}{Department of Physics and Astrophysics, University of Chicago, 5640 S. Ellis Ave, Chicago, IL 60637}

\altaffiltext{3}{Department of Physics and Astronomy, Johns Hopkins University, 3701 San Martin's Drive, Baltimore, MD 21218}

\altaffiltext{4}{Princeton University Observatory, Peyton Hall, Princeton, NJ 08544}

\altaffiltext{5}{Department of Physics and Astronomy, Pennsylvania State University, University Park, PA 16802}

\altaffiltext{6}{Apache Point Observatory, 2001 Apache Point Road, P.O. Box 59, Sunspot, NM 88349-0059}

\altaffiltext{7}{Adler Planetarium and Astronomy Museum, 1300 Lake Shore Drive, Chicago IL 60605}

\authoremail{cwyip@phyast.pitt.edu; ajc@phyast.pitt.edu}         
 
\begin{abstract}

We study 16,707 quasar spectra from the
Sloan Digital Sky Survey (SDSS) (an early version of the First Data Release; DR1) using the
Karhunen-Lo\`eve (KL) transform (or Principal Component Analysis,
PCA). The redshifts of these quasars range from 0.08 to 5.41, the
$i$-band absolute magnitudes from $-30$ to $-22$, and the resulting
restframe wavelengths from 900~\AA \, to 8000~\AA. The quasar
eigenspectra of the full catalog reveal the following: 1st order --- the
mean spectrum; 2nd order --- a host-galaxy component; 3rd order --- the
UV-optical continuum slope; 4th order --- the correlations of Balmer
emission lines. These four eigenspectra account for 82~\% of the total
sample variance. Broad absorption features are found not to be
confined in one particular order but to span a number of higher orders.
We find that the spectral classification of quasars is
redshift and luminosity dependent, as such there does not exist a
compact set (i.e., less than $\approx 10$ modes) of eigenspectra (covering 900~\AA \, to 8000~\AA) 
which can describe most variations (i.e., greater than $\approx 95$~\%) of the
entire catalog. We therefore construct several sets of eigenspectra in
different redshift and luminosity bins. From these eigenspectra we find
that quasar spectra can be classified (by the first two eigenspectra)
into a sequence that is defined by a simple progression in the steepness
of the slope of the continuum.  We also find a dependence on redshift
and luminosity in the eigencoefficients. The dominant redshift effect is
a result of the evolution of the blended \ion{Fe}{2} emission (optical) and the
Balmer continuum (the ``small bump'', $\lambda_{rest} \approx
2000-4000$~\AA).  A luminosity dependence is also present in the
eigencoefficients and is related to the Baldwin effect --- the decrease
of the equivalent width of an emission line with luminosity, which is
detected in Ly$\alpha$, \ion{Si}{4}+\ion{O}{4}$]$, \ion{C}{4}, \ion{He}{2}, 
\ion{C}{3}$]$ and \ion{Mg}{2},
while the effect in \ion{N}{5} seems to be redshift dependent. If we restrict ourselves
to the rest-wavelength regions $1150-2000$~\AA \, and $4000-5500$~\AA, 
the eigenspectra constructed from the wavelength-selected SDSS spectra
are found to agree with the principal components by Francis et~al. (1992)
and the well-known ``Eigenvector-1'' \cite{Boroson92} respectively.
ASCII formatted tables of the eigenspectra are available.

\end{abstract}
 
\keywords{(galaxies:) quasars: general---surveys---methods: statistical---techniques: spectroscopic}
 
\section{Introduction} \label{section:intro}

Quasars (QSOs) serve as tools, in conjunction with studies of the
intergalactic medium, for probing conditions in the early
universe. These studies rely on the fact that the spectra are, to the lowest order, rather 
uniform (e.g., the construction and application of QSO composite spectra). We know,
however, that the spectra do exhibit differences: the spectral slopes, as
well as the line profiles, differ among quasars. In fact, even in a single spectrum, the widths 
of the emission lines can be vastly different. Although these differences may provide
insights for understanding the physical environments in the vicinity of
quasars (by constructing inflow or outflow models for different kinds of elements in the
surroundings), they present substantial challenges when modeling broad and narrow
line regions (BLRs and NLRs). A quantitative understanding of the variation in quasar 
spectra is therefore a necessary and important study.

In the pioneering work by Francis {et~al.} (1992), the authors applied a
Principal Components Analysis (PCA) to 232 quasar spectra (i.e., spectral PCA, in which 
the concerned variables are the observed flux densities in the wavelength bins of a spectrum) from the Large
Bright Quasar Survey (LBQS; Hewett et~al. 1996) and found that the mean spectrum plus the
first two principal components in the rest-wavelength range
$1150-2000$~\AA \, describe the majority of the variation seen in the UV-optical spectra of quasars. 
In this spectral region, the quasars are shown to have a variety of spectral slopes and equivalent
widths, ranging from broad, low-equivalent-width
lines to narrow, high-equivalent width lines, with other spectral
properties also varying along this trend. Furthermore, Boroson and Green
(1992) identified several important parameters in describing quasars and
carried out a PCA on 87 quasars from the Bright Quasar Survey (BQS; Schmidt \& Green 1983) 
in this parameter space (i.e., parameter PCA, in which the variables are the physical 
quantities of interest), from which an anti-correlation was
found between \ion{Fe}{2} (optical, around the H$\beta$ spectral region) and
$[$\ion{O}{3}$]$. (This correlation is widely quoted as ``Eigenvector-1''). More
recently, Shang {et~al.} (2003) considered a wider rest-wavelength range
covering Ly$\alpha$ to H$\alpha$, and constructed eigenspectra from 22
optically selected quasars from the BQS.  Their results
agreed with Boroson and Green's Eigenvector-1, and supported the
speculated anti-correlation between \ion{Fe}{2} (optical) and \ion{Fe}{2} (UV).  The
conclusions of these studies, however, are drawn from small ranges of
redshifts ($1.8 < z < 2.7$; $z < 0.5$; $0.07 < z < 0.4$) respectively.

The SDSS spectroscopic survey has the advantage of a large number of
quasars, and most importantly, a large redshift range.  It provides a
unique opportunity for investigating how quasars differ from one another,
and whether they form a continuous sequence (Francis {et~al.} 1992). In
this paper, we apply the Karhunen-Lo\`eve (KL) transform to study this
problem in the 16,707 quasars from the SDSS.  The primary goals of this
paper are to 1) obtain physical interpretations of the eigenspectra, 
2) determine the effects of redshift and luminosity on the spectra of quasars, 
and 3) study the correlations between broad emission lines and UV-optical continua.
With this  data set, in which $\approx 94$~\% of quasars were discovered by the SDSS, 
our analysis is the most extensive of its kind to date.

We discuss the SDSS quasar sample used in this work in \S~\ref{section:data}, followed
by a review of the KL transform and the gap-correcting procedures in
\S~\ref{section:KL}.  The set of quasar eigenspectra for the whole
sample covering $900-8000$~\AA \, in rest-wavelength are presented in
\S~\ref{section:global}. We quantitatively detect the redshift and
luminosity effects through a commonality analysis of the eigenspectra
sets constructed from quasar subsamples in \S~\ref{section:similar}. The
quasar eigenspectra in several subsamples of different redshifts and luminosities
are shown in \S~\ref{section:zbin}, and we make a comparison between the
KL-reconstructed spectra using either sets of eigenspectra (i.e., the
subsamples versus the global case).  In \S~\ref{section:crossbin}, we
perform a KL transform on cross-redshift and -luminosity bins, from
which evolutionary (\S~\ref{section:evolution}) and luminosity effects
(\S~\ref{section:baldwin}) are found in the quasar spectra. In
\S~\ref{section:class}, we discuss the possible classification of quasar
spectra by invoking the eigencoefficients in these subsamples.
Correlations among the broad emission lines and the local eigenspectra are presented in
\S~\ref{section:linecorr}, including the well-known ``Eigenvector-1''.
\S~\ref{section:conclusion} summarizes and concludes the present work.

\section{Data} \label{section:data}

The sample we use is an early version of the First Data Release (DR1; Abazajian {et~al.} 2003) 
quasar catalog \cite{Schneider03} from the Sloan Digital Sky Survey (SDSS; York {et~al.} 2000), 
which contains 16,707 quasar spectra and was created on the 9th of July, 2003.
The official DR1 quasar catalog includes slightly more objects (16,713) and was created on 
the 28th of August, 2003. All  spectra in our sample are cataloged in the official DR1 quasar catalog
except one: \mbox{SDSS J150322.94+600311.3} (i.e., there are 7 DR1 QSOs not included in our sample).
The SDSS operates a CCD camera \cite{Gunn98} on a 2.5~m telescope located at Apache Point Observatory, 
New Mexico. Images in five broad optical bands (with filters $u, g, r, i$ and $z$; Fukugita {et~al.} 
1996) are being obtained over $\approx 10,000$ 
deg$^{2}$ of the high Galactic latitude sky. The astrometric calibration is described
in Pier et~al. (2003). The photometric system is described in Smith et~al. (2002)
while the photometric monitoring is described in Hogg et~al. (2001). 
The details of the target selection, the spectroscopic reduction
and the catalog format are discussed by Schneider {et~al.} (2003) and  references therein.
About $64$~\% of the quasar candidates in our sample are chosen based
on their locations in the multi-dimensional SDSS color-space \cite{Richards02a}, while 
$\approx 22$~\% are targeted solely by the Serendipity module. The remaining QSOs are primarily targeted
as FIRST sources, ROSAT sources, stars or galaxies.
All quasars in the DR1 catalog have absolute magnitudes ($M_{i}$) brighter than $-22.0$,
where $M_{i}$ are calculated using cosmological parameters $H_0 = 70$ 
km s$^{-1}$ Mpc$^{-1}$, $\Omega_{M}=0.3$ and $\Omega_{\Lambda}=0.7$; and that the UV-optical spectra
can be approximated by a power-law $(f_\nu \propto \nu^{\alpha_{\nu}})$ with the frequency index 
$\alpha_{\nu} = -0.5$ \cite{VandenBerk01}. The absolute magnitudes in five bands are corrected 
for Galactic extinction using the dust maps of Schlegel, Finkbeiner \& Davis (1998). 
Quasar targets are assigned to the 3\arcsec \, diameter fibers for spectroscopic
observations (the tiling process; Blanton et~al. (2003)). Spectroscopic observations are discussed in detail 
by York {et~al.} (2000); Castander {et~al.} (2001); Stoughton {et~al.} (2002) and Schneider {et~al.} (2002). 
The SDSS Spectroscopic Pipeline, among other procedures, removes skylines and atmospheric absorption bands, and
calibrates the wavelengths and the fluxes. The signal-to-noise ratios generally meet the
requirement of $(S/N)^{2}$ of 15 per spectroscopic pixel \cite{Stoughton02}.
The resultant spectra cover $3800-9200$~\AA \, in the observed frame with a spectral resolution
of $1800-2100$.  At least one prominent line in each spectrum in the DR1 quasar catalog is of  
full-width-at-half-maximum (FWHM) $\ge 1000$ km s$^{-1}$.
Type~II quasars and BL~Lacs are not included in the DR1 quasar catalog. 

All of the 16,707 quasars are included in our present analysis, including quasars with 
broad absorption lines (BALQSOs). To perform the KL transforms, the spectra are shifted to their
restframes, and linearly rebinned to a spectral resolution $\approx 1800/(1+z_{min})$,
with $z_{min}$ being the lowest redshift of the whole sample (\S~\ref{section:global})
or of the subsamples of different $(M_{i}, z)$-bins (defined in \S~\ref{section:zbin}). Skylines
and bad pixels due to artifacts are removed and fixed with the gap-correction procedure
discussed in \S~\ref{section:KL}.

Unless otherwise specified, in this paper we present every quasar spectrum as flux 
densities in the observed frame and wavelengths in the restframe for the
convenience of visual inspection. Following the
convention of the SDSS, wavelengths are expressed in vacuum values. 

\section{KL Transform and Gap Correction} \label{section:KL}

The Karhunen-Lo\`eve transform (or Principal Component Analysis, PCA)
is a powerful technique used in classification and dimensional
reduction of massive data sets. In astronomy, its applications in studies of multi-variate 
distributions have been discussed in detail (Efstathiou \& Fall 1984; Murtagh \& Heck 1987). 
The basic idea in applying the KL transforms in studying the spectral
energy distributions is to derive from them  a lower dimensional set of 
{\it eigenspectra} \cite{Con95}, from which the essential physical properties are represented and hence a 
compression of data can be achieved. Each spectrum can be thought of as an axis in a multi-dimensional 
hyperspace, $f_{\lambda_{k} i}$, which denotes the flux density per unit wavelength at the $k$-th 
wavelength in the $i$-th quasar spectrum.

For the moment, we assume that there are no gaps in each spectrum; we will discuss the ways we deal 
with missing data later.  From the set of spectra we construct the correlation matrix
\begin{equation}
  C_{{\lambda_{k}}{\lambda_{l}}} = \hat{f}_{\lambda_{k} i} \hat{f}_{i \lambda_{l}} \ ,
\end{equation}

\noindent
where the summation is from $i=1$ to the total number of spectra, $N$, and
$\hat{f}_{\lambda_{k} i}$ is the normalized $i$-th spectrum, defined for a given $i$ as
\begin{equation}
  \hat{f}_{\lambda_{k}} = {{f_{\lambda_{k}}}\over{\sqrt{{\sum_{\lambda_{k}}^{} f_{\lambda_{k}}  f_{\lambda_{k}}}}}}\ .
\end{equation}

The eigenspectra are obtained by finding a matrix, $U$, such that
\begin{equation}
 U^{T} C U = \Lambda \ ,
\end{equation}

\noindent
where $\Lambda$ is the diagonal matrix containing the eigenvalues of
the correlation matrix. $U$ is thus a matrix whose $i$-th column
consists of the $i$-th eigenspectrum ${e}_{i {\lambda_{k}}}$.  We
solve this eigenvalue problem by using Singular Value Decomposition.

The observed spectra are projected onto the eigenspectra to obtain
the eigencoefficients. In these projections, every wavelength bin in each
spectrum is weighted by the error associated with that particular wavelength bin, 
$\sigma_{\lambda}$, such that the weights are given by 
$w_{\lambda}=1/{\sigma_{\lambda}}^2$. The observed spectra can be decomposed, with 
no error, as follows
\begin{equation}
 {f}_{\lambda_{k}} = \sum_{i=1}^{M} a_{i} e_{i {\lambda_{k}}}\ ,
\end{equation}

\noindent
where $M$ is the total number of eigenspectra, and $a_{i}$ are the expansion coefficients 
(or the {\it eigencoefficients}) of the $i$-th order. It is straightforward
to see that, if the number of spectra is greater than the number of wavelength bins, 
$M$ equals the total number of wavelength bins in the spectrum.

An assumption that the spectra are without any gaps was made previously.
In reality, however, there are several reasons for gaps to exist:
different rest-wavelength coverage, the removal of skylines, bad pixels
on the CCD chips all leave gaps at different restframe wavelengths for
each spectrum. All can contribute to incomplete spectra. The idea behind
the gap-correction process is to reconstruct the missing regions in the spectrum using
its principal components. The first application of this method to
analyze galaxy spectra is due to Connolly \& Szalay~(1999), which
expands on a formalism developed by Everson \& Sirovich (1994) for dealing
with two-dimensional images. Initially, we fix the missing
data by some means, for example, linear-interpolation.  A set of
eigenspectra are then constructed from the gap-repaired quasar
spectra. Afterward, the gaps in the original spectra are corrected with
the linear combination of the KL eigenspectra. The whole process is
iterated until the set of eigenspectra converges. From our previous work
on the SDSS galaxies \cite{Yip04}, the eigenspectra set converges both
as a function of iteration steps in the gap-repairing process 
and the number of input spectra.

To measure the commonality between two sets of eigenspectra (i.e., how
alike they are), two subspaces $E$ and $F$ are formed respectively for the two sets. 
The sum of the projection operators of each
subspace is calculated as follows
\begin{eqnarray}
 \mathbf{E} = \sum_{\epsilon} |\epsilon> <\epsilon| \ ,  
\end{eqnarray} 

\noindent 
where $|\epsilon>$ are the basis vectors which span the space $E$ (see, for example, Merzbacher 1970). 
A basis vector is an eigenspectrum if $E$ is considered to be a set of eigenspectra. 
If the two subspaces are in common, we have
\begin{eqnarray}
 Tr(\mathbf{EFE}) & = & D \ ,
\end{eqnarray} 

\noindent 
where $Tr(\mathbf{EFE})$ is the trace of the products of the projection operators,
and $D$ is the (common) dimension of both subspaces. The two subspaces are disjoint if the trace quantity is
zero, which hence serves as a quantitative measure for
the similarity between two arbitrary subspaces of the same dimensionality.

\section{Global QSO Eigenspectra} \label{section:global}

Models of accretion on black holes and  scenarios for the formations of \ion{Fe}{2}-blends
often predict  relationships between the UV and optical quasar spectral properties (for example, the strong  
anti-correlation between the ``small bump'' and the optical \ion{Fe}{2} blends was suggested by 
Netzer and Wills 1983). Using our sample with 16,707 quasar spectra, we construct a set of eigenspectra 
covering 900~\AA \, to 8000~\AA \, in the restframe. For each quasar spectrum, the spectral regions
without the SDSS spectroscopic data are approximated by the
linear combinations of the calculated eigenspectra by the gap-correction
procedure described in \S~\ref{section:KL}. A quantitative assessment of
this procedure on quasar spectra is discussed in detail in Appendix~\ref{appendix:gapcorr}. To determine
the number of iterations needed for this gap-correcting procedure, we
calculate the commonality between the two subspaces spanned by the
eigenspectra in one iteration step and those in the next step. For the
subspace spanned by the first two modes, the convergence rate is fast
and it requires about three iterations at most
to converge. Including higher-order components, in this case the first
100 modes, the subspace takes about 10 iteration steps to converge. In
this work, all eigenspectra are corrected for the missing pixels 
with 10 iteration steps. The gaps in each spectrum
are corrected for using the first 100 eigenspectra during the iteration.

\begin{figure}
\epsscale{0.8}
\plotone{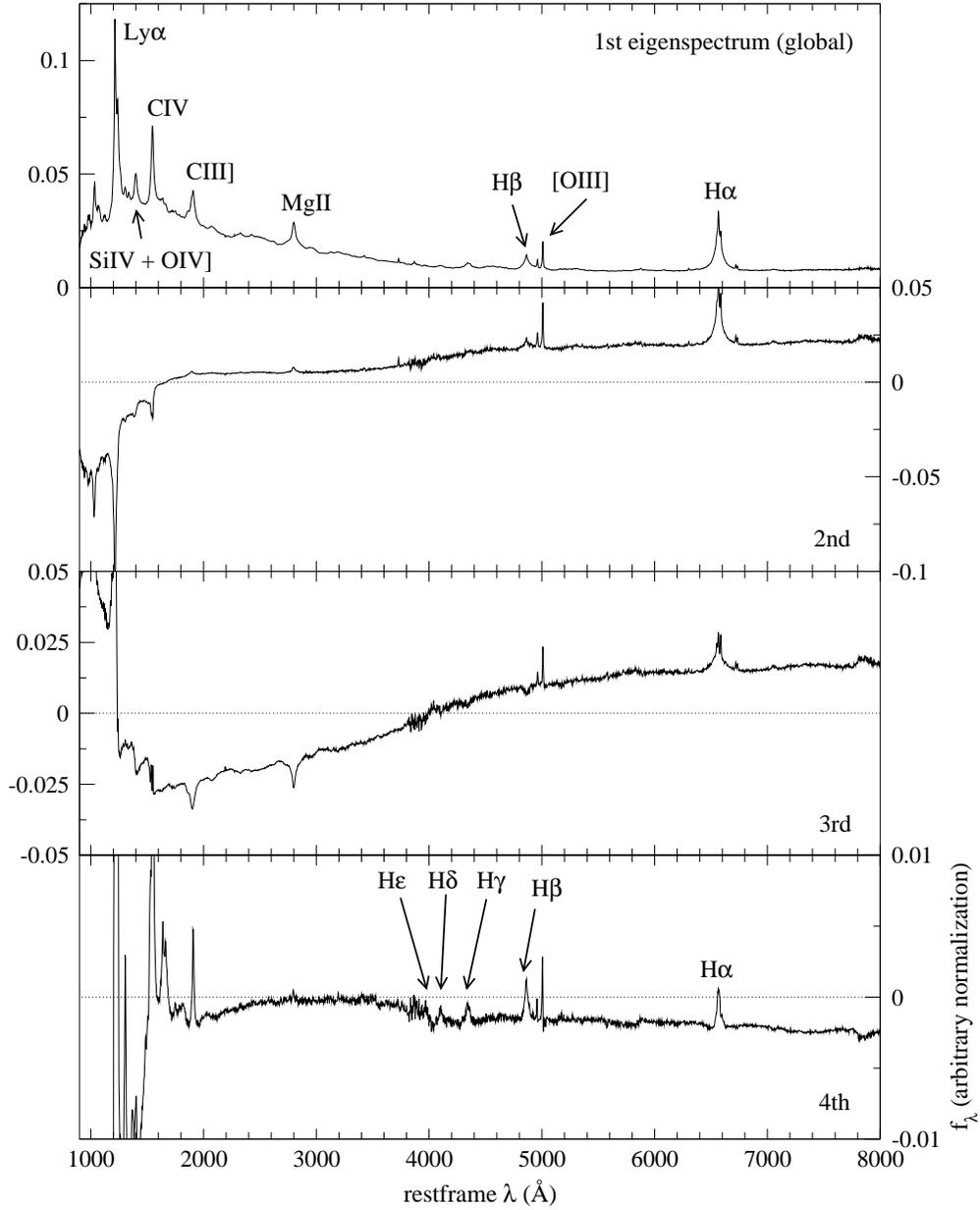}
\figcaption[fig1.eps]{The first 4 eigenspectra of 16,707 SDSS quasars, 
in the rest wavelengths $900-8000$~\AA. Prominent emission lines are indicated.~\label{fig:global_eigenspec}}
\end{figure}

The partial sums of weights (i.e., accumulative weights, where the weights
are the eigenvalues of the correlation matrix) in different orders of the 
global eigenspectra are shown in Table~\ref{tab:weights_all}.  The first eigenspectrum accounts
for about 0.56 of the total sample variance and the first 10 modes 
account for $\approx 0.92$. To account for $\approx 0.99$ of the total
sample variance, about $50-60$ modes are required.
The first four eigenspectra are shown in Figure~\ref{fig:global_eigenspec}, 
and their physical attributes will be discussed below.

\subsection{First Global Eigenspectrum: Composite Spectrum}

\begin{figure}
\plotone{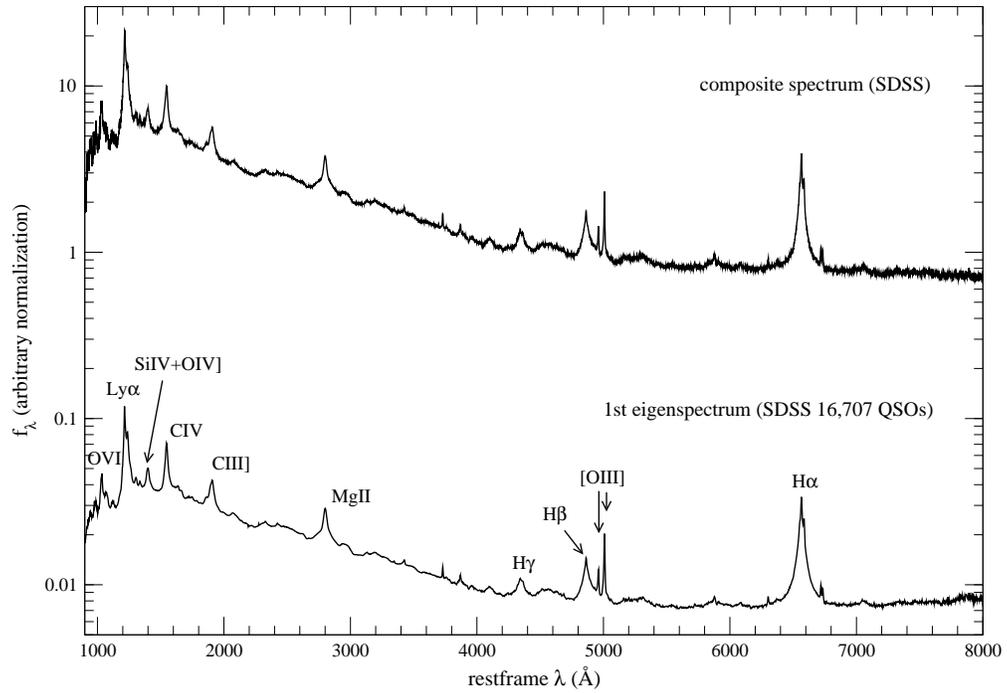}
\figcaption[fig2.eps]{The 1st eigenspectrum of 16,707 SDSS quasars,
 in the rest-wavelengths $900-8000$~\AA. For comparison, the SDSS composite quasar spectrum~\cite{VandenBerk01} 
using over 2200 QSOs is also shown. Prominent emission lines are indicated.~\label{fig:compare_1steigSpec_edrcomposite}}
\end{figure}

The first eigenspectrum (the average spectrum of the data set) reveals the dominant broad emission lines 
that exist in the range of $\lambda_{rest}=900-8000$~\AA. These, presumably Doppler-broadened lines,
are common to most quasar spectra. As can be seen in Figure~\ref{fig:compare_1steigSpec_edrcomposite},
this eigenspectrum exhibits a high degree of similarity
with the median composite spectrum \cite{VandenBerk01} constructed using over 2200 SDSS quasars, 
but with lesser noise at the blue and red ends,
probably due to the larger sample used in this analysis.

\subsection{Second Global Eigenspectrum: Host-Galaxy Component}

\begin{figure}
\plotone{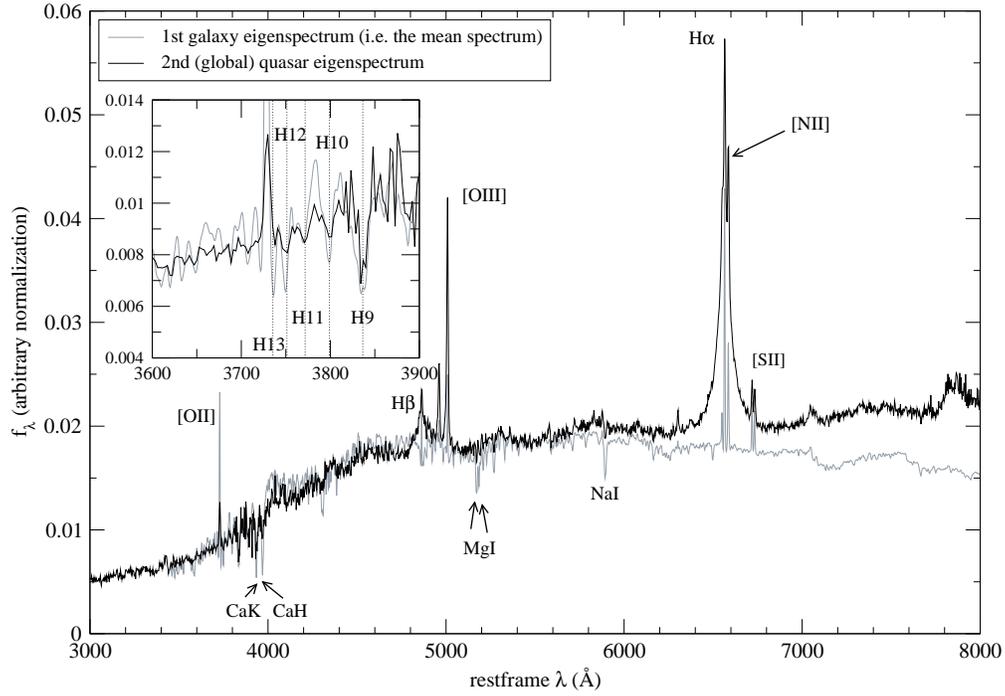}
\figcaption[fig3.eps]{Comparison of the global quasar eigenspectrum of this work with
the 1st eigenspectrum (i.e., the mean spectrum) of the SDSS galaxies ($\approx 170,000$ galaxy spectra,
from Yip {et~al.} 2004) in the rest-wavelengths $3000-8000$~\AA. Not only the major emission lines
and absorption lines noted in the graph are found in both cases, the ``bumps and wiggles'' in the
continua also exhibit similarities. The inset shows the spectral region near the hydrogen absorption lines.~\label{fig:compare_qso_gal}}
\end{figure}

The 2nd eigenspectrum shows a striking similarity in the optical region ($\lambda_{rest} \ge 3500$~\AA)
with the 1st galaxy eigenspectrum (i.e., mean spectrum) from the SDSS
galaxies (of $\approx 170,000$ galaxy spectra; Yip {et~al.}
2004). Figure~\ref{fig:compare_qso_gal} shows a comparison between the
two. Besides the presence of the \xion{Ca}{K} and \xion{Ca}{H} lines and the
Balmer absorption lines as reported previously in the composite quasar
spectrum, the \ion{Mg}{1} triplet 
(which appears to be composed of two lines because of the limited
resolution, i.e., \ion{Mg}{1}$\lambda$5169+$\lambda$5174, and 
\ion{Mg}{1}$\lambda$5185\footnote[1]{The restframe wavelength of the longest wavelength component of 
the \ion{Mg}{1} triplet appears to be redshifted by $\approx 290$~km~s$^{-1}$ relative to the laboratory 
vacuum value of 5185~\AA. This maybe due to the contamination by an unidentified absorption line
redward of \ion{Mg}{1}$\lambda$5185.}) is also seen in this mode. 
The presence of the Balmer absorption lines (see the inset of Figure~\ref{fig:compare_qso_gal}) implies 
the presence of young to intermediate stellar populations near the nuclei (because of the SDSS
3\arcsec \, spectroscopic fiber). The main differences between the
quasar 2nd eigenspectrum and the galaxy mean spectrum lie in the Balmer
lines H$\alpha$ and H$\beta$, which are, as expected,
Doppler-broadened for the QSO spectra. The quasar eigenspectrum also has
a redder continuum, meaning that {\it if} this eigen-component
represents all contributions from the host-galaxies, the galaxies would
be of earlier spectral type than the average spectral type in the SDSS Main galaxy sample.

Our ability to detect significant host-galaxy features in this eigenspectrum triggers an important
application, that is, the removal of the host-galaxy contributions from
the quasar spectra.  The properties the host-galaxies of quasars have
recently attracted interest (e.g., Bahcall et~al. 1997, McLure {et~al.} 1999, McLure {et~al.}
2000, Nolan {et~al.} 2001, Hamann {et~al.} 2003), mainly because of
their obvious relationship with the quasars they harbor and the
probable co-evolution that happens between them. Therefore, the
evolution of massive galaxies, which are believed to be at one
time active quasar hosts (see Hamann \& Ferland 1999), can also be probed.

On the other hand, narrow emission lines in active galactic nuclei
(AGNs) have been considered less useful than broad emission lines as
diagnostic tools, because AGNs with prominent narrow lines have low
luminosities (see, for example, the discussion in Chapter~10 of Krolik 1999), 
in which case contributions from the host galaxies may
affect both the continuum and the lines, obscuring their true
appearances. Hence, the removal of host-galaxy components can
potentially fix the narrow emission lines and reveal their true physical
nature. Preliminary results (Vanden Berk et~al. 2004, in prep.)
show that it is possible to remove the galaxy continuum in the
lower-redshift quasars in the SDSS sample.  Related issues such as the
effects on the broad and narrow emission lines from such a removal
procedure are beyond the scope of this paper and are currently being
studied.

The second mode also shows slight anti-correlations 
between major broad emission lines which exist in $\lambda_{rest}$ smaller and larger than  
$\approx 2000$~\AA \, (see Figure~\ref{fig:global_eigenspec}).

\subsection{Third Global Eigenspectrum: UV-Optical Continuum Slope}\label{section:globalslope}

The change of the continuum slope, with a zero-crossing (i.e., a node) at
around 3990~\AA, dominates this global eigenspectrum. The optical
continuum appears to be galaxy-like, but not as much as the 2nd global
eigenspectrum. For example, in this component the [\ion{O}{2}]$\lambda$3728 is
missing, and the nebular lines are generally weaker. The node at
$\approx 4000$~\AA \, is in partial agreement with the 2nd principal
component of 18 low-redshift ($z < 0.4$; BALQSOs {\it excluded}) quasar spectra \cite{Shang03},
which showed the UV-optical continuum variation (except the node is at
$\approx 2600$~\AA). This particular wavelength (4000~\AA) marks the
modulation of the slope between the UV and the optical regions.  One
related effect is the ``ultra-violet excess'', describing the abrupt
rise of quasar flux densities from about 4000~\AA \, to 3500~\AA. This
observed excess flux was suggested to be due to the Balmer continuum
\cite{Malkan82}, as there seem to be no other mechanisms which can
explain this wavelength coincidence. In Malkan \& Sargent's work, an exact wavelength for this onset
was not clear. The node at $\approx 4000$~\AA \, can serve the purpose of defining that wavelength. Other
possible physical reasons for the modulations between the UV and optical
continua are the intrinsic change in the quasar continuum (e.g., due to
intrinsic dust-reddening) and the stellar light from the host galaxy.
There is also a second node located in Ly$\alpha$ showing an anti-correlation between
the continua blueward and redward of the Ly$\alpha$. Since the number
of quasars with spectroscopic measurements in the vicinity of Ly$\alpha$ is much smaller than
those with measurements in the UV-optical regions that are 
redward of Ly$\alpha$, the significance of this anti-correlation
is less than that of the UV-optical continuum variation in this eigenspectrum.

\subsection{Fourth Global Eigenspectrum: Correlations of Balmer Emission Lines}

This mode shows the correlations of broad emission lines, namely, Ly$\alpha$, \ion{C}{4}, \ion{Si}{4}+\ion{O}{4}$]$, \ion{C}{3}$]$, 
\ion{Mg}{2}, $[$\ion{O}{3}$]\lambda$5008 and also the Balmer emission lines H$\alpha$, H$\beta$,
H$\gamma$, H$\delta$ and H$\epsilon$.  These are in partial agreement
with the 3rd eigenspectrum of Shang {et~al.}, in which emission lines
\ion{C}{3}$]$, \ion{Mg}{2}, H$\alpha$, H$\beta$ are found to be involved. It
seems natural that these Balmer lines are correlated, as presumably they
are formed coherently by some photo-ionization processes.  However, it
is not known why they appear in this low-order mode.  The fact
that \ion{C}{3}$]$ and H$\beta$ vary similarly was seen previously
\cite{Wills00}, and it was suggested that H$\beta$ and \ion{C}{3}$]$ may
arise from the same optically-thick disk.

\subsection{Higher Orders} \label{section:globalhighorder}

By construction, subsequent higher-order eigenspectra show more nodes,
causing small modulations of the continuum slope.
They also show broad absorption line features. Since quasars
with BALs are not the dominating populations in our sample (there are 224 broad absorption line
quasars in the 3814 quasars from the SDSS EDR quasar catalog,
Reichard {et~al.} 2003),  their signatures preferentially show up at higher
orders in this global set of eigenspectra.  
The BAL components are not confined to only one particular mode,
but span a number of orders. To investigate the effects of BALQSOs on the global eigenspectra,
our approach is to perform the KL transform on our original sample (including BALQSOs) 
and on the same sample but with the BALQSOs excluded, and make
a comparison between them. There are 682 BALQSOs (with balnicity index $> 0$) found in our sample
according to the BALQSO catalog for the SDSS spectra by Trump et~al.~(private communication). 

\begin{figure}
\plotone{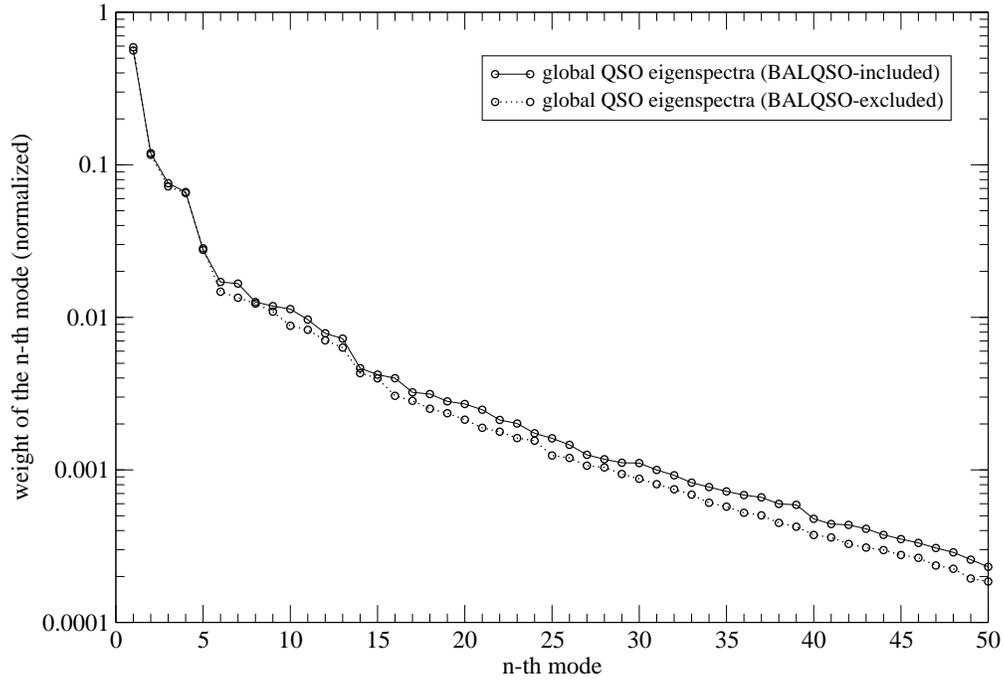}
\figcaption[fig4.eps]{Comparison of the weights at different orders
 between the BALQSO-included global eigenspectra (i.e., the 16,707 QSOs; solid curve) 
and the BALQSO-excluded global eigenspectra 
(dotted curve). Since the BALQSO-included global eigenspectra contain 
 information for features of both typical quasars and broad absorption lines, the weight of each mode is
larger than that of the BALQSO-excluded eigenspectra.~\label{fig:weight_global_BAL_BALexcluded}}
\end{figure}

Figure~\ref{fig:weight_global_BAL_BALexcluded} compares the weights at different orders
between the BALQSO-included and the BALQSO-excluded global eigenspectra.
Since the BALQSO-included global eigenspectra contain 
information describing both the non-BALQSOs and the BALQSOs, the weight of each mode is
larger than that of the BALQSO-excluded eigenspectra. That is, the BALQSO-excluded eigenspectra set 
is more compact. The magnitude of this offset, however, is small and is apparent only  
after the 5-th order, which is consistent with the fact that the BALQSOs form a minority population 
(about $4$~\%). This difference is seen to extend to  higher orders, implying that 
the features describing the BALQSOs span a number of higher-order eigenspectra
and are not confined to only one particular mode. 

\begin{figure}
\plotone{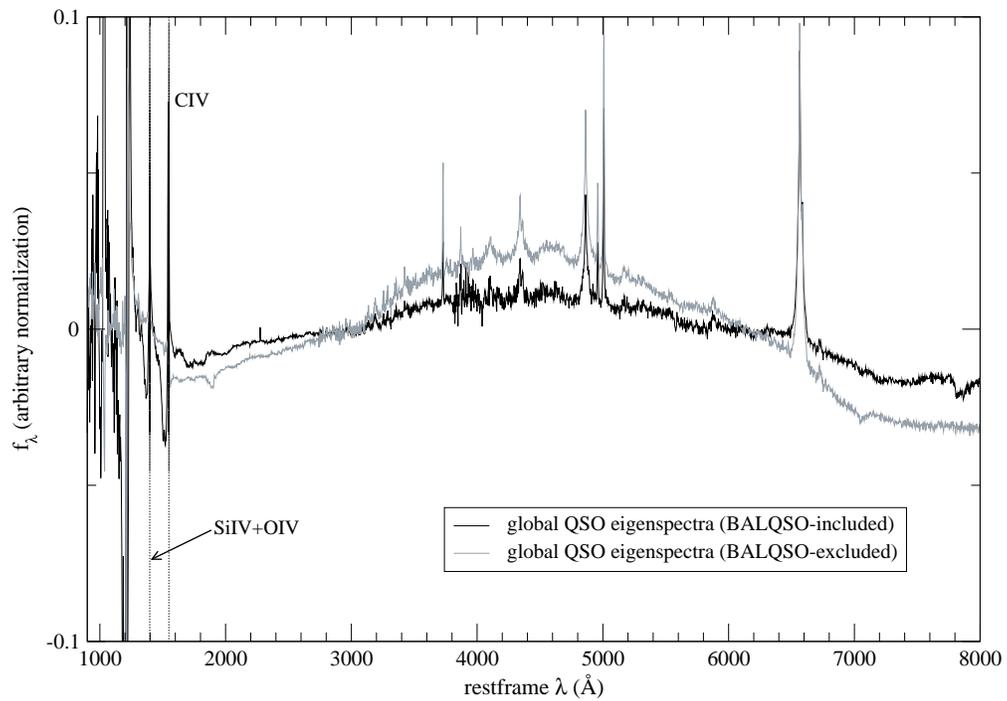}
\figcaption[fig5.eps]{Comparison between the 6th global eigenspectra from 
the BALQSO-included sample and the BALQSO-excluded one. Absorption features in \ion{Si}{4}+\ion{O}{4}$]$ and 
\ion{C}{4} exist in the first case and is missing in the latter, 
as expected.~\label{fig:comp_all_allcutBAL_6thmode}}
\end{figure}

A comparison of the 6th global eigenspectrum between the BALQSO included and excluded samples
is shown in Figure~\ref{fig:comp_all_allcutBAL_6thmode}. Absorption features (in this case,
in \ion{Si}{4}+\ion{O}{4}$]$ and \ion{C}{4}) are found
in the first set of eigenspectra but are missing in the latter. 
We have to note that the discrepancies in the spectral features of these two sets of
eigenspectra attributed to the weight differences are not only confined to
the existence or non-existence of BAL absorption troughs as shown here, as the difference
in the normalizations between the two can in general also yield different eigenspectra sets.
We will leave the discussion of the reconstruction of the 
BALQSO spectra using eigenspectra till \S~\ref{section:BAL} .

\subsection{A Non-Unique Set of Eigenspectra: Commonality Analysis} \label{section:similar}

To study the possible evolution and luminosity effects in the quasar
spectra, our first step is to investigate whether the set of
eigenspectra of a given order derived from quasar spectra in different
redshift and luminosity ranges differ. The trace quantity mentioned in
\S~\ref{section:KL} is adopted for these quantitative comparisons.

\begin{figure}
\plotone{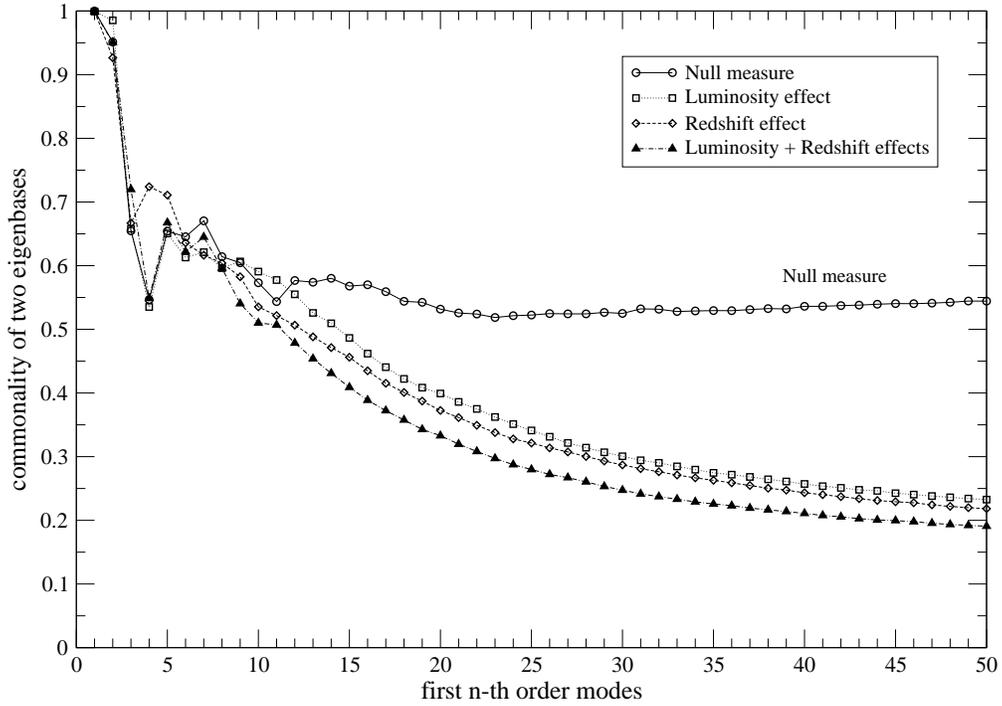}
\figcaption[fig6.eps]{Commonality of two subsamples which are designed to
be different from each other mainly due to noise and intrinsic variations (solid line with circles), 
the additional luminosity effect (dotted line with squares), 
redshift effect (dashed line with diamonds), and both (dot-dashed line with filled triangles); 
plotted against the common dimension of the eigenbases 
(which equals the number of modes constructing each eigenbasis). The commonality departs from unity
and progressively drops below the null measure. This means that the two eigenbases under consideration
are more disjoint from each other when higher the orders are included. ~\label{fig:common_subsample}}
\end{figure}

As a null measure, two subsamples are chosen with approximately the same
redshift and luminosity distributions, such that any differences in the
two sets of eigenspectra would be due to noise and the intrinsic
variability of the quasars. We fix the rest-wavelengths of this study to
be $2000-4000$~\AA, and require a full rest-wavelength coverage of
the input quasars; redshifts are limited to 0.9 to 1.1. One subsample
contains 472 objects (Subsample 1) and the other subsample, 236 objects
(Subsample 2). Subsample 2 is, by construction, a subset of the original
472 objects. The reason behind this construction is to ensure a high
commonality of the two sets of resultant eigenspectra.  They both have
luminosities from $-24$ to $-25$, and the actual distributions of redshifts
and luminosities are similar.  The line on the top in
Figure~\ref{fig:common_subsample} shows the commonality of these two
subsamples as we increase the number of eigenspectra forming the
subspace.  As higher orders of eigenspectra are included in the
subspaces, the commonality drops, meaning that the two subspaces become
more disjoint. As mentioned above, this disjoint behavior is mainly due
to the noise and the intrinsic variability among quasars, both are
unlikely to be completely eliminated. At about 20 modes and higher, the
commonality levels off, which implies that the eigenspectra mainly
contain noise.

With this null measure in place, the differences of our test subsamples
are further relaxed to include luminosity effects alone (Subsamples 1
and 3, see Table~\ref{tab:subsample}), redshift effects alone
(Subsamples 3 and 4), and lastly, both effects combined (Subsamples 1
and 4).  The commonalities of these subsamples are overlaid in
Figure~\ref{fig:common_subsample}.  The first modes constructed in all
these subsamples, including the null measure, are always very similar to
each other (more than 99~\% similar). This shows that a single mean
spectrum can be constructed across the whole redshift coverage, which
was presumed to be true in many previous constructions of quasar
composite spectra. The validity of construction of the mean spectrum in
a given sample may seem trivial, but it is not if we take into account
the possibility that the quasar population may evolve at different
cosmic epochs.

Similar to the null measure, as higher orders are included in the
subspaces, the eigenspectra subspaces become more disjoint.  In
addition, the commonalities in these condition-relaxed cases actually
drop {\it below} the null measure for orders of modes higher than $\approx 10$. 
Therefore, the eigenspectra of the same order but derived from
quasars of different redshifts and luminosities describe different
spectral features.  In addition, our results show that both luminosity and
evolution effects have detectable influences on the resultant sets of
eigenspectra, very much to the same degree (in terms of commonality).
In the case of the combined effects, the commonality drops to the lowest
value among all cases, as expected.

The actual redshift and luminosity effects found in the quasar spectra
will be presented in Sections~\ref{section:evolution} and
\ref{section:baldwin}. We learn from this analysis that
there does not exist a unique set of KL eigenspectra across the whole
redshift range, with the number of modes equal or smaller than approximately
10. The implications are twofold. On one hand, the classification of
quasar spectra, in the context of the eigenspectra approach, has to be
redshift and luminosity dependent. In other words, the {\it weights} of
different modes are in general different when quasars of different
redshifts and luminosities are projected onto the same set of
eigenspectra.  So, eigenspectra derived from quasars of a particular
redshift and luminosity range in general do not {\it predict} quasar
spectra of other redshifts and luminosities. On the other hand, the
existence of the redshift and luminosity effects in our sample can be
probed quantitatively by analyzing the eigenspectra subspaces.

\section{QSO Eigenspectra in  $(M_{i}, z)$-bins} \label{section:zbin}

KL transforms are performed on subsamples with different redshift and luminosity ranges,
that allow us to explicitly discriminate the possible luminosity effects on the spectra 
from any evolution effects, and vice versa\footnote[2]{Since the K-correction of our
sample is calculated in the SDSS assuming a spectral index $\alpha_{\nu} = -0.5$, so in principle
a color dependence is present in any redshift trend found.}.  
The constructions of these bins are based on
requiring that the maximum gap fraction among the quasars, that is, the
wavelength region without the SDSS data, is smaller than 50~\% of the the
total spectral region we use when applying the KL transforms. The total
spectral region, by construction, is approximately equal to the largest
common rest-wavelengths of all the quasars in that particular bin. We
find that constraining the gap fraction to be a maximum of 50~\% improves
the accuracy of the gap-correcting procedure for most quasars (see Appendix~\ref{appendix:gapcorr}
for further explanation). 
As a result, five divisions are made in the whole redshift range $0.08 < z < 5.13$ (where
the quasars of redshifts larger than 5.13 are discarded to satisfy the constraint of
50~\% minimum wavelength-coverage in all related luminosity bins), 
and four in the whole luminosity range $M_{i}=(-30, -22)$. 
These correspond to {\bf ZBIN~1} to {\bf 5} and the $M_{i}$ bins {\bf A} to {\bf
D} for the redshift and luminosity subsamples respectively. 
In the following, we denote each subsample in a given luminosity and redshift range, for example, 
the bin {\bf A4}. Such divisions are by no means unique and can be constructed according to
one's own purposes, but we find that important issues such as the
correlation between continua and emission lines remain unchanged as we
construct bins with slightly different coverages in redshift, in
luminosity and in the total rest-wavelength range. 
The actual rest-wavelength range and the number of spectra in each bin are shown
in Table~\ref{tab:cuts}, which also lists the fractions of QSOs in each bin that are
targeted either in the quasar color-space \cite{Richards02a} or solely by the Serendipity module.  
While the majority of the quasars from most of the bins are targeted by 
using the multi-dimensional color-space, in which the derived eigenspectra are expected to be 
dominated the intrinsic quasar properties,
there is one bin ({\bf C4}) in which most quasars are targeted by the Serendipity module.
In principle, the eigenspectra in the latter case will represent the properties of the 
serendipitous objects and lack a well motivated color distribution.

\begin{figure}
\epsscale{0.78}
\plotone{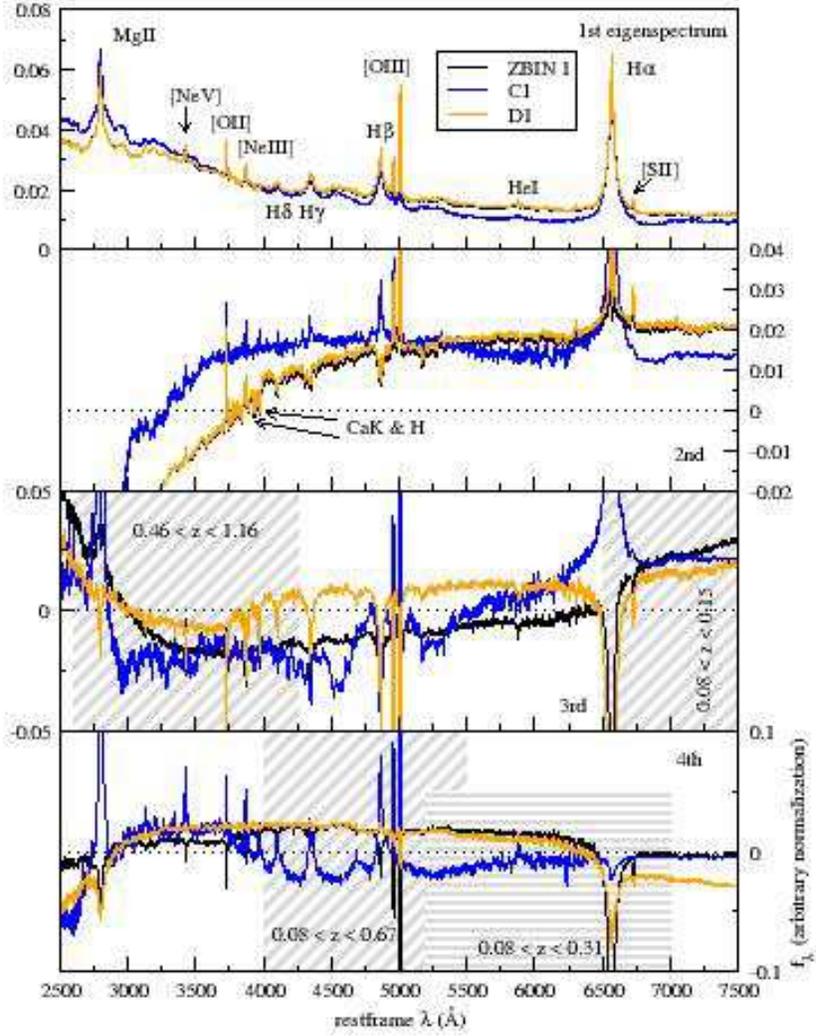}
\figcaption[fig7.eps]{The first 4 orders of ($M_{i}, z$)-binned eigenspectra 
for the subsample {\bf ZBIN~1}. The shaded areas
correspond to the local spectral properties as found in the  2nd local eigenspectra sets 
 using wavelength-selected quasar spectra. (Lowered Resolution) \label{fig:ZBIN1_eigenspec}}
\end{figure}

\begin{figure}
\epsscale{0.78}
\plotone{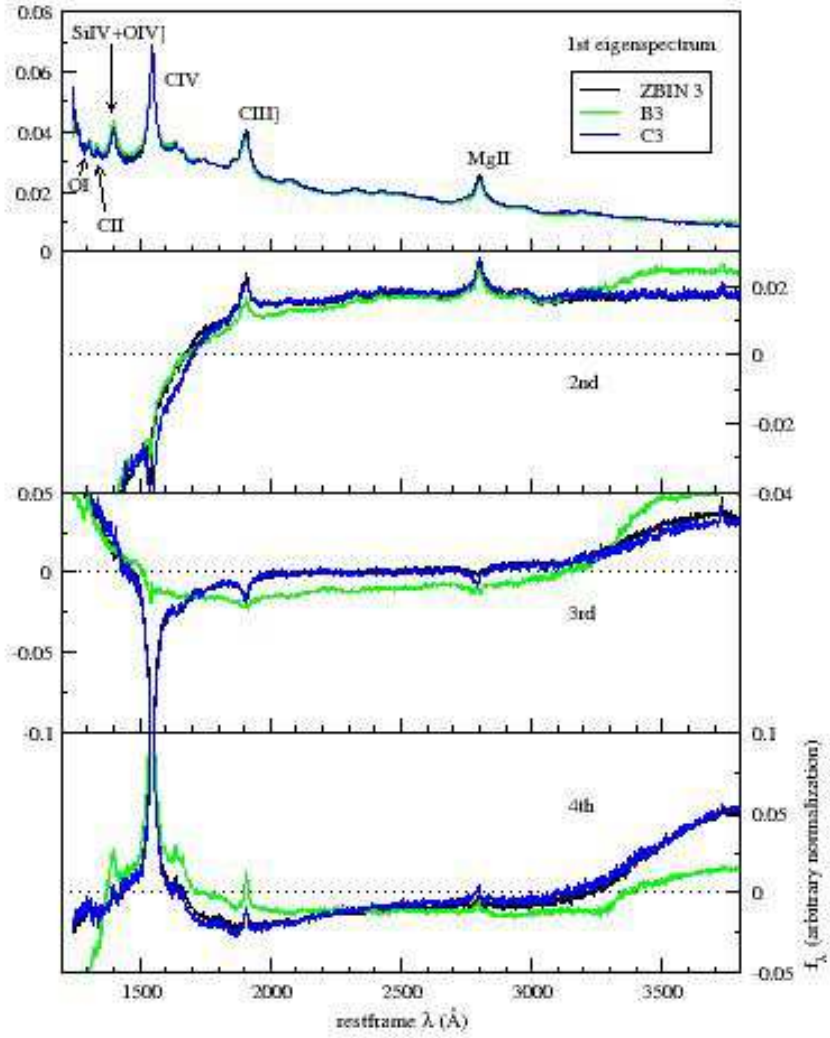}
\figcaption[fig9.eps]{The first 4 orders of ($M_{i}, z$)-binned eigenspectra for 
the subsample {\bf ZBIN~3}. (Lowered Resolution)
\label{fig:ZBIN3_eigenspec}} 

\end{figure}

\begin{figure}
\epsscale{0.78}
\plotone{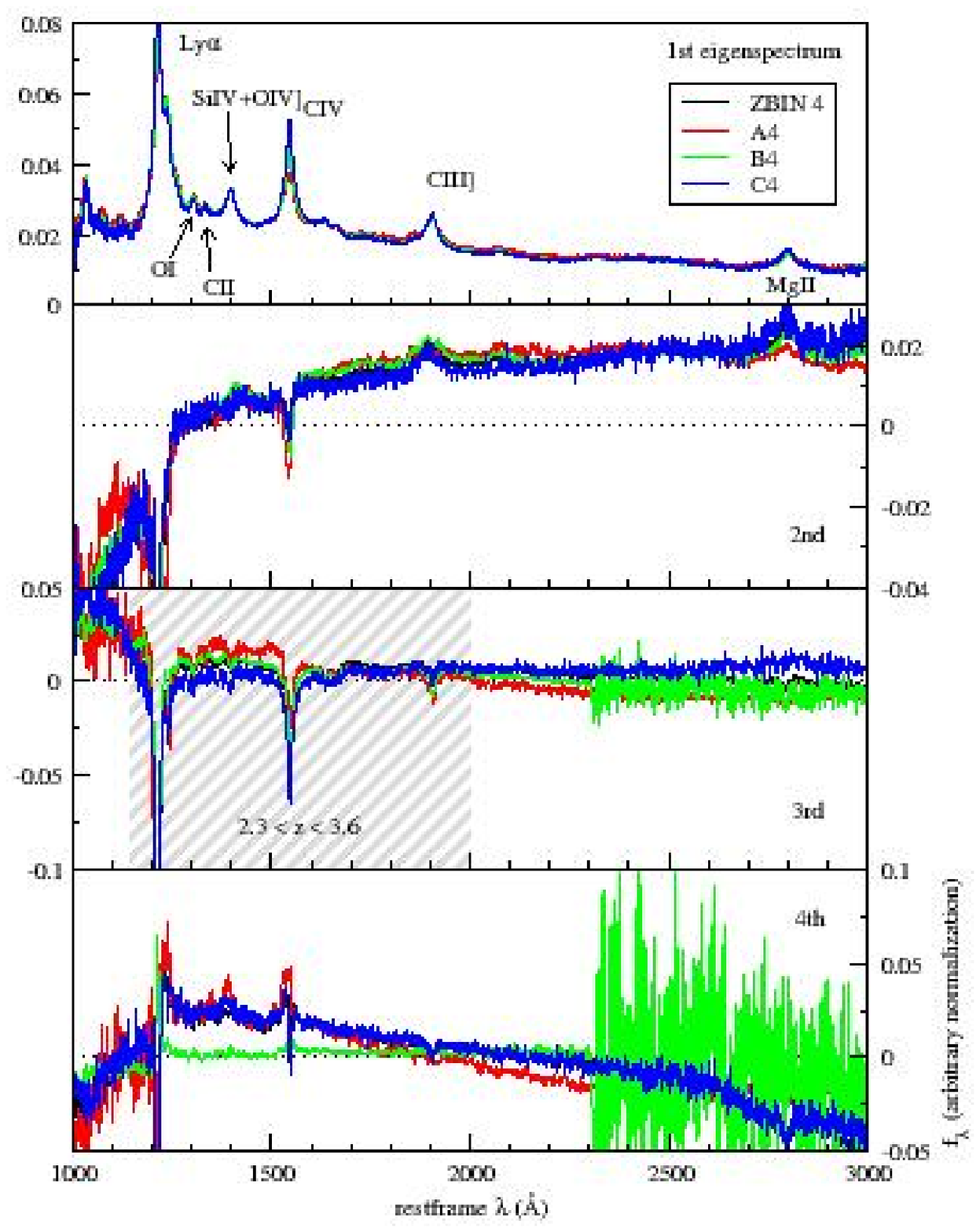}
\figcaption[fig10.eps]{The first 4 orders of ($M_{i}, z$)-binned  eigenspectra for
 the subsample {\bf ZBIN~4}. The meaning of the shaded spectral region is 
explained in the caption of Figure~\ref{fig:ZBIN1_eigenspec} and \S~\ref{section:localinMizbin}.
(Lowered Resolution) \label{fig:ZBIN4_eigenspec}} 
\end{figure}

\begin{figure}
\epsscale{0.78}
\plotone{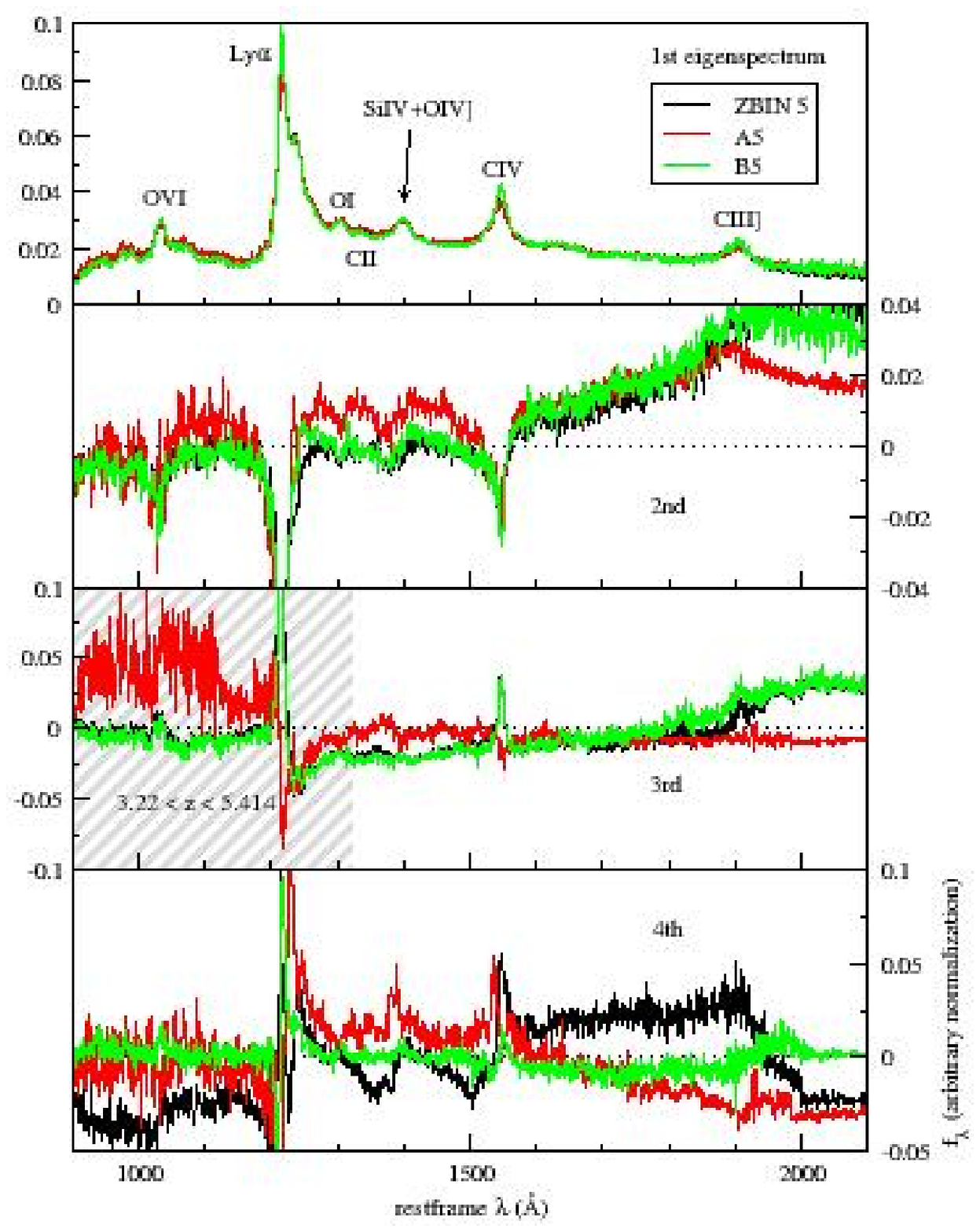}
\figcaption[fig11.eps]{The first 4 orders of eigenspectra of the subsample {\bf ZBIN~5}.
The meaning of the shaded spectral region is 
explained in the caption of Figure~\ref{fig:ZBIN1_eigenspec} and \S~\ref{section:localinMizbin}.
(Lowered Resolution) \label{fig:ZBIN5_eigenspec}} 
\end{figure}

In general for all  $(M_{i}, z)$-bins, the first 10 modes or less are required to account for  
more than 92~\% of the variances of the corresponding spectra sets (Table~\ref{tab:weights}).
In the iterated calculation of the $(M_{i}, z)$-binned eigenspectra, the first 50 modes are used in the 
gap correction. The first 4 orders of eigenspectra 
of each ($M_{i}, z$)-bin are shown in Figures~\ref{fig:ZBIN1_eigenspec} $-$ \ref{fig:ZBIN5_eigenspec},
arranged in 5 different redshift ranges. 
In each figure, eigenspectra of different luminosities are plotted
along with the ones which are constructed by combining all luminosities (shown in black curves).
By  visual inspection, the eigenspectra in different orders show diverse
properties for each $(M_{i}, z)$-bin. In the following, properties associated with
different orders are extracted by considering {\it all} $(M_{i}, z)$-bins generally. Eigenspectra
which are distinct from the average population will be discussed separately. 

\subsection{First $(M_{i},z)$-Eigenspectra: Composite Spectra}

As in the global case, the lowest-order eigenspectra are simply the mean of the quasars in the given subsamples. 
For every redshift bin, the first eigenspectrum shows approximately a power-law shape (either
a single or broken power-law), with prominent broad emission lines. 
Different luminosity bins show differences in the overall spectral slopes to  various
degrees. In every redshift range, the spectra of higher-luminosity quasars are bluer 
than their lower luminosity counterparts. For example, {\bf C1} 
(Figure~\ref{fig:ZBIN1_eigenspec}; $M_{i}= -26 - -24$) shows a 
harder spectral slope blueward  of $\approx 4000$~\AA \,  than that of {\bf D1} ($M_{i} = -24 - -22$). 
However, for the higher redshift ($z = 2.06 - 5.13$) quasars, e.g., in {\bf ZBIN~4} (Figure~\ref{fig:ZBIN4_eigenspec}) 
and {\bf 5} (Figure~\ref{fig:ZBIN5_eigenspec}), the difference in spectral slope seems to be confined 
mainly to changes in the flux densities blueward of Ly${\alpha}$.

\subsection{Second $(M_{i},z)$-Eigenspectra: Spectral Slopes}\label{section:slope}

The 2nd mode in every $(M_{i},z)$-bin has one node at a particular wavelength. This implies 
that the linear-combination of the first 2 modes changes the spectral slope. This 
is similar to the galaxy spectral  classification by the KL approach \cite{Con95}, 
in which the first two eigenspectra give the spectral shape. 

For the lowest redshift bin ({\bf ZBIN~1}; Figure~\ref{fig:ZBIN1_eigenspec}), the node of the second 
eigenspectrum occurs at about 3850~\AA \, for the lower luminosity QSOs ({\bf D1}), but at $\approx 3300$~\AA \, 
for the higher luminosity ones ({\bf C1}). Possible physical reasons underlying the modulation of the UV-optical 
 slopes were discussed previously in \S~\ref{section:globalslope}. Interestingly,  
the luminosity averaged 2nd eigenspectrum (black curve) in this redshift range also shows galactic features 
(as found for the 2nd global eigenspectrum). The continuum redward of $\approx 4000$~\AA \, 
is very similar to that in galaxies of earlier-type. Absorption lines \xion{Ca}{K} and \xion{Ca}{H}, 
and the Balmer absorption lines H~9, H~10, H~11 and H~12 are seen in the lower-luminosity bin {\bf D} 
(and are not present in 
the higher-luminosity bin {\bf C}, hence a luminosity dependent effect is implied). 

\subsection{Third $(M_{i},z)$-Eigenspectra:  Anti-correlation between \ion{Fe}{2} (UV) and 
optical continuum around H$\beta$} \label{section:FeAntiCorr}

\begin{figure}
\epsscale{0.78}
\plotone{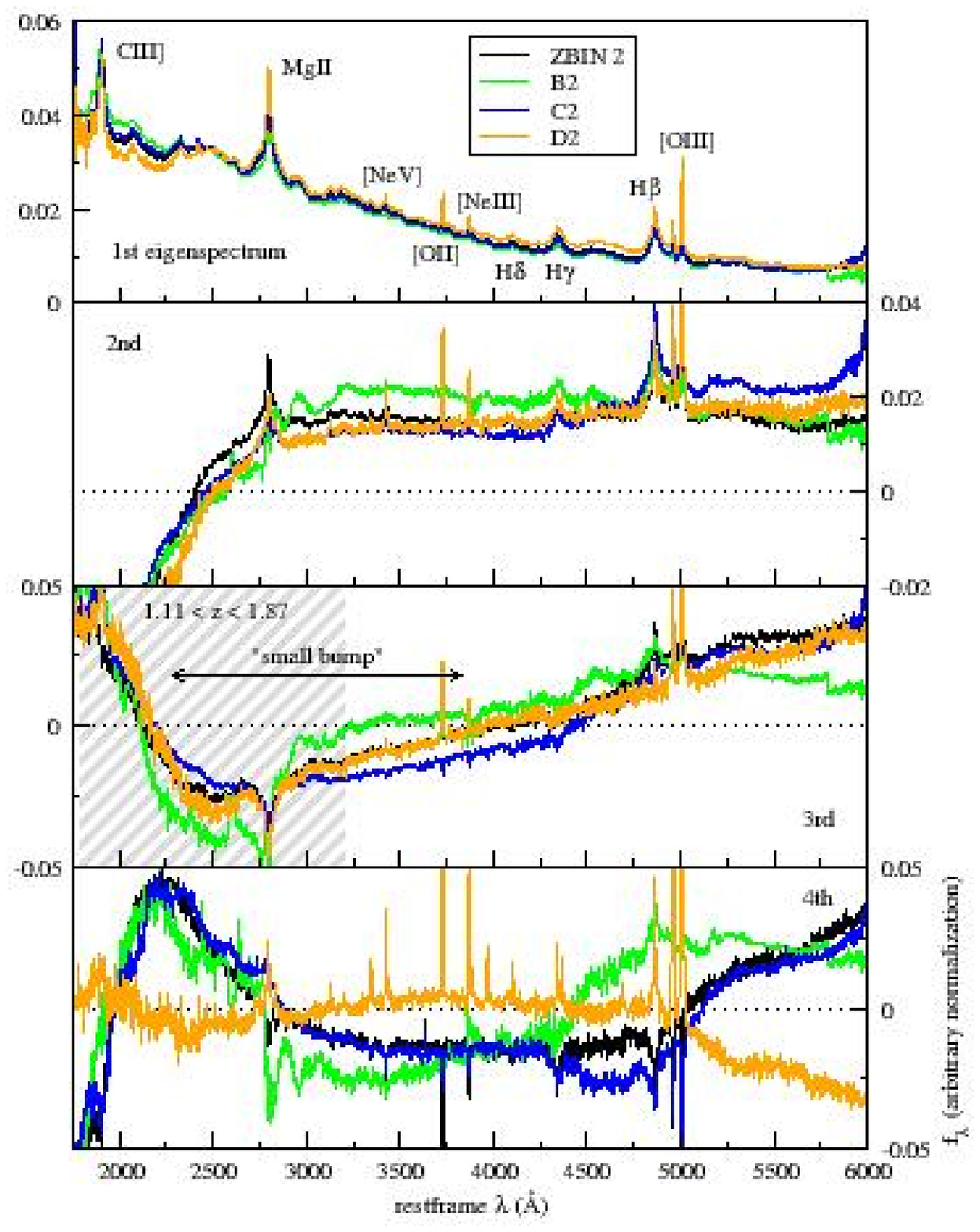}
\figcaption[fig8.eps]{The first 4 orders of ($M_{i}, z$)-binned 
eigenspectra for the subsample {\bf ZBIN~2}. The meaning of the shaded spectral region is 
explained in the caption of Figure~\ref{fig:ZBIN1_eigenspec} and 
\S~\ref{section:localinMizbin}. (Lowered Resolution) \label{fig:ZBIN2_eigenspec}} 
\end{figure}

In addition to the finer-modulation of the continuum slope provided by the 3rd eigenspectrum compared with the
2nd mode, in the redshift range $0.53< z <1.16$ ({\bf ZBIN~2}; Figure~\ref{fig:ZBIN2_eigenspec}), 
averaging over all luminosities, this mode shows a strong
anti-correlation between the quasi-continuum in the \ion{Fe}{2} (UV) regions around \ion{Mg}{2} 
(the ``small bump'', with its estimated location indicated in the 3rd eigenspectrum in Figure~\ref{fig:ZBIN2_eigenspec})
and the continuum in the vicinity of H$\beta$. Around the H$\beta$ emission, the continuum is blended with the
\ion{Fe}{2} optical blends, the H$\delta$, H$\gamma$ and [\ion{O}{3}] lines. The wavelength bounds are found to be 
$\approx 2120-4040$~\AA  \, for the \ion{Fe}{2} ultraviolet blends and $\approx$ 4050~\AA \, upward 
(to $\approx$ 6000~\AA, which is the maximum wavelength of this redshift bin) for the optical 
continuum around H$\beta$. This appears to support the calculations 
that strong \ion{Fe}{2} optical emissions require a high optical depth in the resonance 
transitions of the \ion{Fe}{2} (UV) \cite{Netzer83, Shang03}, hence a decrease in the strength of the latter.
The actual wavelengths of the nodes bounding the \ion{Fe}{2} (UV) region are shown in
 Figure~\ref{fig:ZBIN2_eigenspec}. For brighter quasars ({\bf B2}), the small bump is smaller  
($\approx 2120-3280$~\AA) than that found in fainter QSOs.

\subsection{Reconstructing BALQSOs with $(M_{i}, z)$-eigenspectra} \label{section:BAL}

\begin{figure}
\plotone{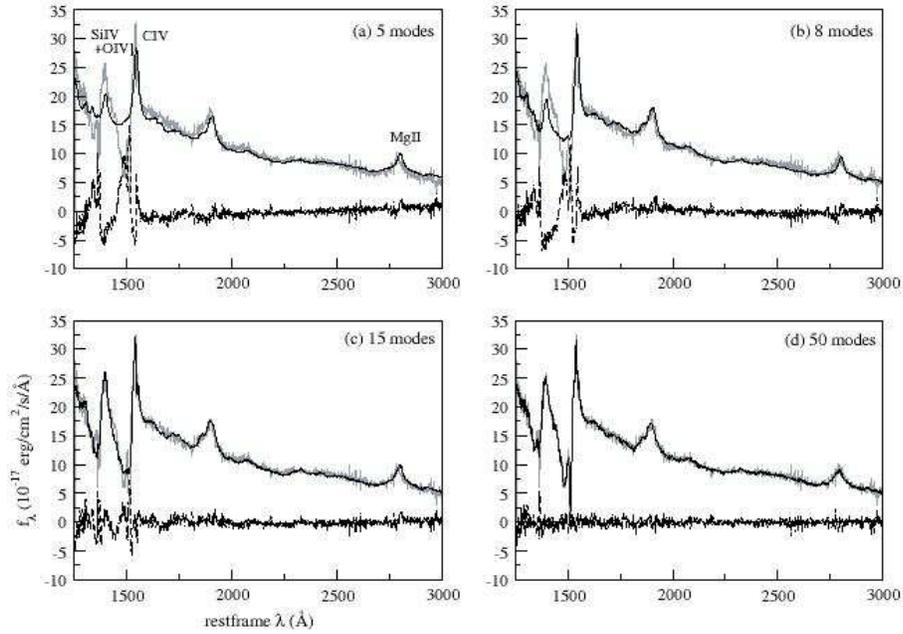}
\figcaption[fig12.eps]{KL-reconstructed spectra (black solid) of a quasar spectrum (gray) with broad
absorption features in \ion{C}{4} and \ion{Si}{4} (\mbox{SDSS J110041.19+003631.9}, classified as HiBAL according to 
Reichard et~al. 2003) using the first (a) 5 modes (b) 8 modes (c) 15 modes (d) 50 modes, where the eigenspectra are 
constructed from the ($M_{i}, z$)-bin {\bf B3}. The bottom dashed curves are the residuals. The broad absorption features
span a number of higher-order modes. The observed spectrum is not smoothed. (Lowered Resolution) 
\label{fig:recon_BAL_B3_277_51908_437}}
\end{figure}

To examine the intrinsic broad absorption line features in the $(M_{i}, z)$-binned eigenspectra, 
we study the reconstructed spectra using different numbers of eigenspectra.
Figure~\ref{fig:recon_BAL_B3_277_51908_437} shows one of the EDR BAL quasars (Reichard et~al. 2003) found in the
bin {\bf B3}, and its reconstructed-spectra using different numbers of eigenspectra. 
This HiBAL (defined as having high-ionization broad absorption troughs such as \ion{C}{4}) 
quasar is chosen for its relatively large absorption trough in \ion{C}{4} for visual clarity. 
The findings in the following are nonetheless general.
The first few modes ($\lsim 8$ for this spectrum) are found to fit mainly the continuum,
excluding the BAL troughs. With the addition of higher-order modes the
intrinsic absorption features (in this case, in the emission lines \ion{C}{4} and \ion{Si}{4}) 
are gradually recovered.
Some intrinsic absorption features are found to require $\approx 50$ modes for accurate description,
as was found in the global eigenspectra (\S~\ref{section:globalhighorder}).
We should note that in the reconstructions using different numbers of modes;
the {\it same} normalization constant is adopted (meaning the eigencoefficients are normalized 
to $\sum_{m=1}^{50} a_{m}^{2} = 1$).
Clearly, a different normalization constant in the case of reconstructions using fewer modes
(e.g., Figure~\ref{fig:recon_BAL_B3_277_51908_437}a) will further improve the fitting in the least-squares sense.

While the fact that a large number of modes are required to reconstruct the absorption troughs 
probably suggests a non-compact set of KL eigenspectra (referring to those defined in this work)
for classifying BAL quasars, the appropriate truncation of the expansion at some order of eigenspectra 
in the reconstruction process will likely lead to an {\it un-absorbed} continuum, invaluable to many applications. 
The proof of the validity of such a truncation will require detailed future analyses.
One method is to construct a set of eigenspectra using only the known BAL quasars in the sample
and to make comparisons between that and our current sets of eigenspectra. 
By comparing the different orders of both sets of eigenspectra we may be able to recover the BAL physics.
We expect that this separate set of BALQSO-eigenspectra  will 
likely reduce the number of modes in the reconstruction, which is desirable from
the point of view of classification.

\subsection{KL-reconstructed Spectra}\label{section:KLrecon}

\begin{figure}
\plotone{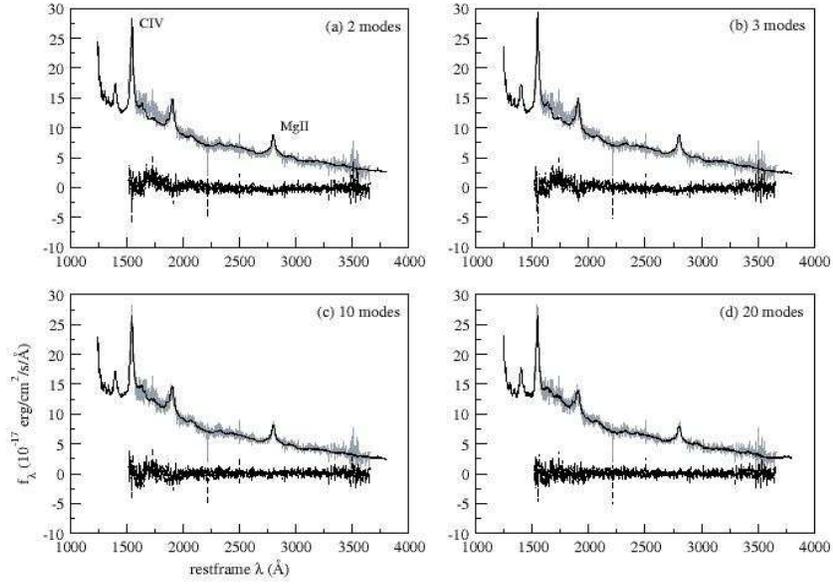}
\figcaption[fig13.eps]{KL-reconstructed spectra (black solid) of an example quasar spectrum (gray) 
(\mbox{SDSS J015214.54+131532.0}) using  the first (a) 2 modes (b) 3 modes (c) 10 modes (d) 20 modes, 
where the eigenspectra are constructed from the ($M_{i}, z$)-bin {\bf C3}. The bottom dashed curves are the residuals.
The observed spectrum is not smoothed. (Lowered Resolution) \label{fig:recon_CZBIN3_spec500}}
\end{figure}

\begin{figure}
\plotone{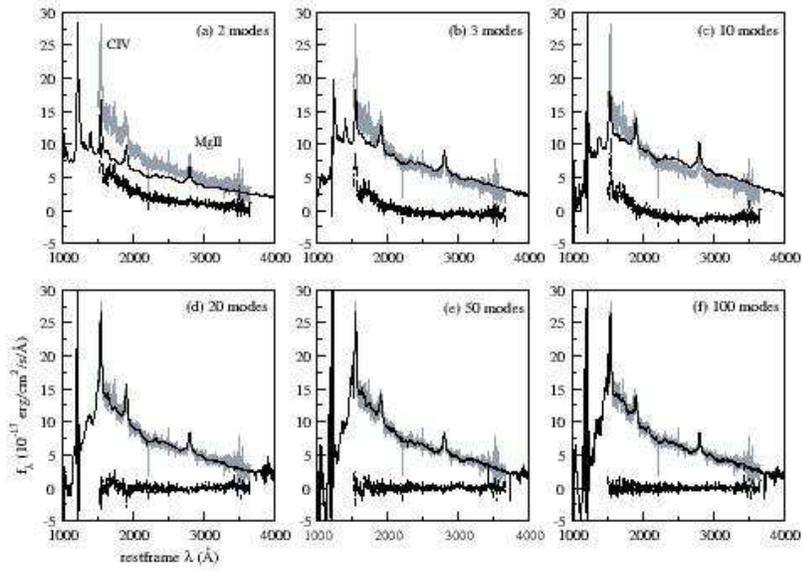}
\figcaption[fig14.eps]{KL-reconstructed spectra (black solid) of the same quasar spectrum (gray) as in 
Figure~\ref{fig:recon_CZBIN3_spec500} using the first (a) 2 modes (b) 3 modes (c) 10 modes (d) 20 modes
(e) 50 modes and (f) 100 modes of the global eigenspectra. 
The bottom dashed curves are the residuals, and the observed spectrum is not smoothed. (Lowered Resolution)
\label{fig:recon_all_spec2017}} 
\end{figure}

Reconstructions of a typical non-BAL quasar spectrum are shown in
Figure~\ref{fig:recon_CZBIN3_spec500}, using from (a) 2 to (d) 20 orders of eigenspectra. 
This particular quasar is in the $(M_{i},z)$-bin {\bf C3}. The bottom curve in each sub-figure shows 
the residuals from the original spectrum. The first 10 modes are
sufficient for a good reconstruction. The reconstructions of the same quasar spectrum but using the global set of
eigenspectra are shown in Figure~\ref{fig:recon_all_spec2017}, from (a) 2 modes to (f) 100 modes.
To obtain the same kind of accuracy, more
eigenspectra are needed in the global case; in this case about 50 modes. 
This is not surprising as the global eigenspectra must account for the intrinsic
variations in the quasar spectra as well as any redshift or luminosity evolutions.

There are, therefore, two major factors we should consider when adopting a global set of quasar eigenspectra 
for KL-reconstruction and classification of quasar (instead of redshift and luminosity dependent sets). 
First, we need to understand and interpret about $10^{2}$ global eigenspectra. This is significantly larger than found
for galaxies (2 modes are needed to assign a type to 
a galaxy spectrum according to Connolly {et~al.}~1995). This is a manifestation of the larger variations in the 
quasar spectra. Second, the ``extrapolated'' spectral region, $\lambda_{rest} < 1520$~\AA\,, in 
Figure~\ref{fig:recon_all_spec2017}  (which is the rest-wavelength region without spectral data) 
show an unphysical reconstruction even when 
100 modes are used, although this number of modes can accurately reconstruct 
the spectral region with data. This agrees with the commonality analysis in 
\S~\ref{section:similar}, that there are evolutionary and luminosity effects in the QSOs in our sample.
As such, eigenspectra derived in a particular redshift and luminosity range are in general not identical to 
those derived in another range. 

The accuracy of the extrapolation 
in the no-data region using the KL-eigenspectra remains an open question for the $(M_{i}, z)$-bins.
It will be an interesting follow-up project to confront the repaired spectral region with observational data,
which ideally cover the rest-wavelength regions where the SDSS does not. For example, UV spectroscopic
observations using the Hubble Space Telescope. 

\section{Evolutionary and Luminosity Effects} \label{section:crossbin}

\subsection{Cross-Redshift and -Luminosity bins Projection}

To study evolution in quasar spectra with the eigenspectra, we must ensure 
that the eigencoefficients reflect the same physics independent of redshift. We 
know however that the eigenspectra change as a function of redshift
(see \S~\ref{section:similar}). To overcome this difficulty, 
and knowing that the overlap spectral region between the two sets of 
eigenspectra in any pair of adjacent redshift bins 
is larger than the common wavelength region ($2124-2486$~\AA) for the full redshift interval, 
we study the differential evolution 
(in redshift) of the quasars by projecting the observed spectra at higher redshift 
 onto the eigenspectra from the adjacent bin of lower redshift. In this way, 
the eigencoefficients can be compared directly from one redshift bin to the next.

Without the loss of generality, we project the observed quasar spectra in the higher redshift bin 
(or dimmer quasars for the cross-luminosity projection) onto
the eigenspectra which are derived in the adjacent lower-redshift one 
(or brighter quasars for the cross-luminosity projection). 
For example, $spec({\bf B3})$ (i.e., the spectra
in the $(M_{i}, z)$-bin {\bf B3}) are projected onto $\{$e$({\bf B2})\}$ (the set of eigenspectra from the 
 $(M_{i}, z)$-bin {\bf B2}), and similarly for the different luminosity bins but the same redshift bin.
From that, we can derive the relationship between the eigencoefficients and redshift (or 
luminosity).

\subsection{Evolution of the Small Bump}\label{section:evolution}

\begin{figure}
\plotone{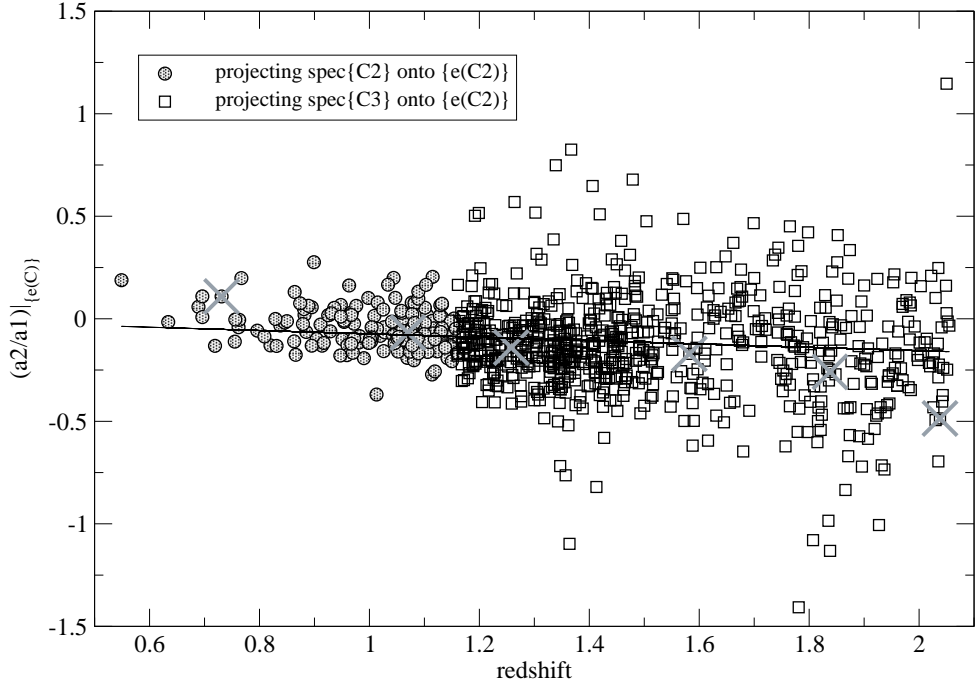}
\figcaption[fig15.eps]{Dependence of $a_{2}/a_{1}$ on redshift, of quasars
within the redshift range $0.53-2.06$, and at a fixed luminosity $M_{i}=-25.5 \pm 0.1$.
The straight line is the regression of the data points: $(a_2/a_1)_{\{e({\bf C2})\}}= - 0.0820 z + 0.0083$. 
The subscript ${\{e({\bf C2})\}}$ denotes that the eigenspectra are constructed from the subsample ${\bf C2}$.
The crosses mark the observed spectra illustrated in Figure~\ref{fig:extraC32_a2_a1_realspec.new}. 
\label{fig:extraC32_a2_a1_redshift}} 
\end{figure}

\begin{figure}
\plotone{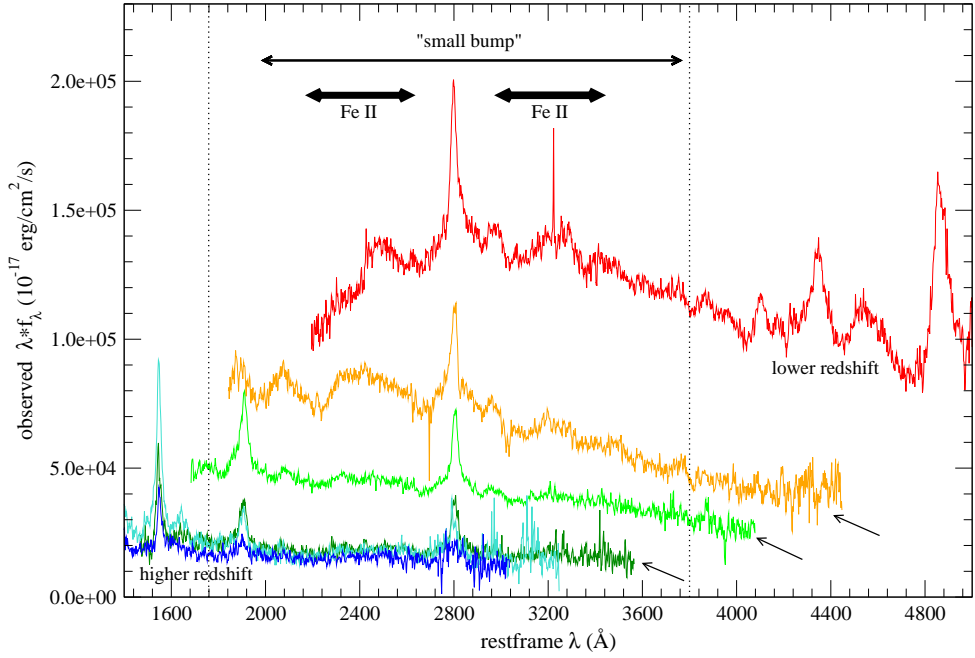}
\figcaption[fig16.eps]{The observed quasar spectra located along the eigencoefficient-redshift 
relation in Figure~\ref{fig:extraC32_a2_a1_redshift} in the original flux scales.  
The evolution of the small bump is evident, in that it is more prominent in the lower-redshift 
quasar spectra. The vertical dotted lines marked the common wavelength region on which
the KL cross-projection is performed. The heavy arrows
mark the wavelength regions in which the \ion{Fe}{2} emissions are typically found. 
The spectra are smoothed with a FWHM $=3$~\AA \, Gaussian smoothing
function for easier visualization. (The QSOs are, from the lowest to highest redshifts,
\mbox{SDSS J173052.71+602516.6}, \mbox{SDSS J015352.65-092010.7}, \mbox{SDSS J090934.26+552944.1}, 
\mbox{SDSS J033801.88+002718.8}, \mbox{SDSS J012858.45+152647.4}, \mbox{SDSS J021552.00-092310.3}.) 
\label{fig:extraC32_a2_a1_realspec.new}} 
\end{figure}

\begin{figure}
\plotone{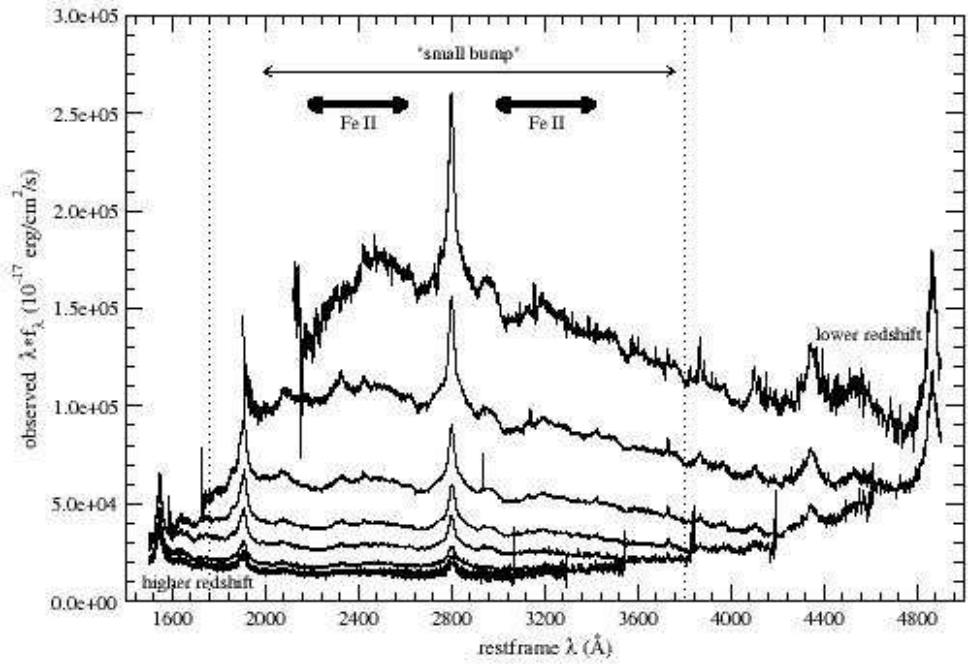}
\figcaption[fig17.eps]{The mean spectra along the redshift trend in Figure~\ref{fig:extraC32_a2_a1_redshift},
from $z = 0.6$ to $2.2$ with a bin width in redshift $dz = 0.2$. 
The observed spectra are not smoothed in the calculations of
the mean spectra. (Lowered Resolution) \label{fig:extraC32_a2_a1_compositeSpec}} 
\end{figure}

The most obvious evolutionary feature is
 the small bump present in the spectra at around $\lambda_{rest} \approx 2000$~\AA \, to $4000$~\AA. 
This feature is mainly composed of blended \ion{Fe}{2} emissions ($\approx 2000-3000$~\AA, Wills {et~al.} 1985) 
and the Balmer continuum ($\approx 2500-3800$~\AA). When we project quasar spectra of redshifts $1.16-2.06$ 
(i.e., $spec({\bf C3})$) onto eigenspectra constructed from quasars of  redshifts $0.53-1.16$ 
(i.e., \{$e$({\bf C2})\}), the coefficients from the second eigenspectrum show 
a clear trend with redshift, as shown in Figure~\ref{fig:extraC32_a2_a1_redshift}. 
In this figure, only those quasars with $M_{i} = -25.5 \pm 0.1$ are chosen (900 objects), as such
the redshift trend does not primarily depend on the absolute luminosities of the quasars. 
To understand this relation {\it observed} spectra are selected along the regression line 
in Figure~\ref{fig:extraC32_a2_a1_redshift} (with the locations marked by the crosses) 
and are shown in Figure~\ref{fig:extraC32_a2_a1_realspec.new}. 
The two dotted lines mark the bandpass where the cross-redshift projection is
performed. The small bump is found to be present and is prominent in the lower-redshift quasars, whereas 
it is small and may be absent in the higher-redshift ones. The spectra marked by the arrows in 
Figure~\ref{fig:extraC32_a2_a1_realspec.new} lie relatively
close to the regression line. An example of the range of evolution in the small bump as a function of redshift
is shown by the remaining 3 spectra which deviate from the regression line.
The observed evolution is present independent of which of the spectra we consider.
The mean spectra (Figure~\ref{fig:extraC32_a2_a1_compositeSpec}) 
as a function of redshift, constructed using a bin width in redshift ($dz$) of 0.2, show a similar
behavior. Each mean spectrum is calculated by averaging the valid flux densities of all objects in each wavelength
bin. The regression of the eigencoefficient-ratios with redshift (with
outliers of $a_2/a_1 > 1$  removed from the calculation) is 
\begin{equation}
 (a_2/a_1)_{\{e({\bf C2})\}}= - 0.0820 z + 0.0083 \ ,
\end{equation}

\noindent
where the subscript $\{e({\bf C2})\}$ denotes that the eigenspectra are from ${\bf C2}$. The correlation
coefficient ($r$) is calculated to be 0.1206 with a two-tailed
P-value\footnote[3]{The P-value for the t-test is calculated under the hypotheses 
$H_{0}: r = 0$ and $H_{1}: r \ne 0$.} of 0.00027 (the probability that we would see such a
correlation at random under the null hypothesis of $H_{0}: r = 0$), 
as such the correlation is considered to be extremely significant by conventional statistical criteria. 

This redshift dependency can be explained by either the evolution of chemical abundances in the quasar environment 
\cite{Kuhn01}, or an intrinsic change in the continuum itself (which, of course, could
also be due to the change in abundances through indirect photo-ionization processes). Green, Forster \&
Kuraszkiewicz (2001) found in the LBQS that the primary correlations of the strengths of \ion{Fe}{2} emission lines
are probably with redshift; an evolutionary effect is therefore implied.
Kuhn {et~al.} (2001) also supported the evolution of
the small bump region $2200-3000$~\AA \, from high-redshift ($\approx 3-4$) to lower-redshifts ($<0.3$) 
by comparing two QSO subsamples with evolved luminosities. 

As the second mode in the $(M_{i}, z)$-binned eigenspectra describes the change in the spectral slope of the sample,
the above findings support the idea that the Balmer continuum, as a part of the small bump, changes with redshift. 
To further understand this effect, the 3rd eigenspectrum in  {\bf C2} is taken into consideration, 
which presumably describes the iron lines (see \S~\ref{section:FeAntiCorr}).
We find that the third eigencoefficient-ratio $a_3/a_1$ also shows a slight redshift
dependency (not shown) with the regression relation (with
outliers of $a_3/a_1 > 1$ removed from the calculation, resulting in 901 objects)
\begin{equation}
 (a_3/a_1)_{\{e({\bf C2})\}}= 0.0478 z -0.2063
\end{equation}

\noindent
and the correlation coefficient is calculated to be 0.0030 with a two-tailed P-value of 0.93,
which is considered to be not statistically significant. 

While the strength of this effect shown by the two ratios are of similar magnitude (0.0820 versus 0.0478), 
the difference in their correlation coefficients
implies that the sample variation is much greater in the ratio $a_{3}/a_{1}$ than $a_{2}/a_{1}$.
The non-trivial value of the regression slope in the case of $a_{3}/a_{1}$ 
agrees with the change in shape of the observed line profiles in the small bump regions seen in the local 
wavelength level (smaller in width than what is expected in the continuum change) with
redshift. In conclusion, this implies that there exists 
the possibility of an evolution in iron abundances 
but with a larger sample variation compared with that for the continuum change.

To our knowledge, our current analysis is the first one without invoking assumptions
of the continuum level or a particular fitting procedure of the \ion{Fe}{2} blends that
finds an evolution of the small bump; directly from the KL eigencoefficients. 
Because of the large sample size, the conclusion of this work that 
the small bump evolves is drawn from spectrum-to-spectrum variation independent
of the luminosity effect, in contrast to the previous composite spectrum approaches \cite{Thompson99},
in which the authors found that the composite spectra in 
two subsamples with mean redshifts $<z>=3.35$ and $<z>=4.47$,
and that from the Large Bright Quasar Survey of lower redshifts 
($<z> \sim 0.8$) are similar in the vicinity of \ion{Mg}{2} and hence did not suggest the existence of a
 redshift effect. The variation of the small bump with redshift is further 
confirmed with the study of composite quasar spectra of the DR1 data set (Vanden Berk {et~al.}, in preparation). 
At this point we make no attempt to quantitatively define and deblend the
\ion{Fe}{2} optical lines and the Balmer continuum, as that would be beyond the scope of this
paper. It is a well-known and  unsolved problem to identify the true shape
of total flux densities due to  the \ion{Fe}{2} emission lines. This difficulty arises because
there are too many \ion{Fe}{2} lines to model and they form a quasi continuum. 

\subsection{Luminosity Dependence of Broad Emission Lines}\label{section:baldwin}

\begin{figure}
\plotone{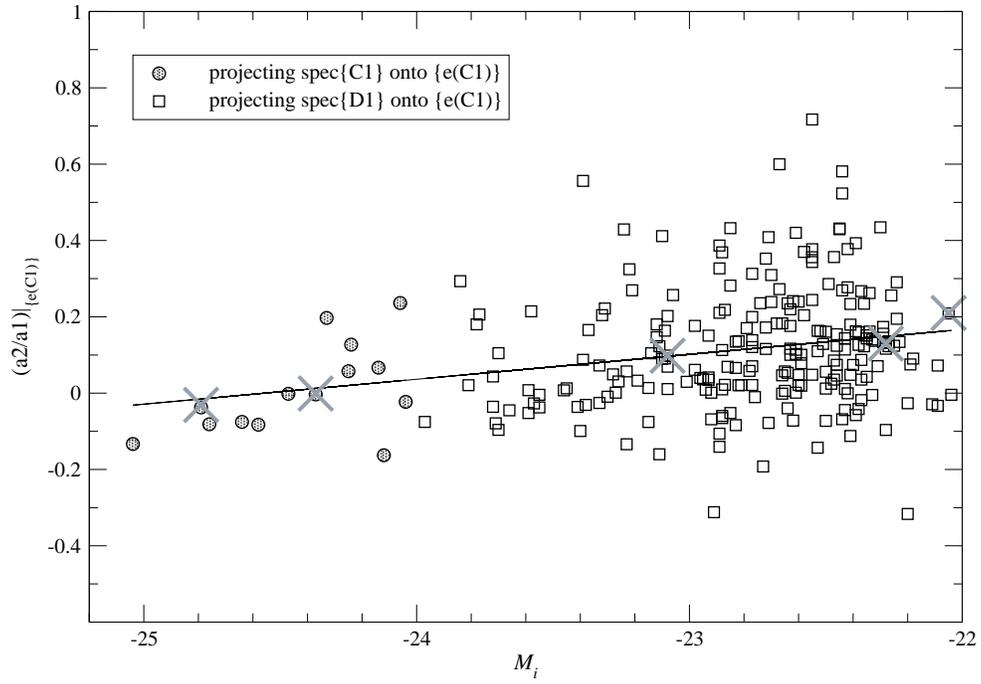}
\figcaption[fig18.eps]{Dependence of $a_{2}/a_{1}$ on the $i$-band
absolute luminosity, of quasars with fixed redshifts ($z=0.4 \pm 0.02$).
The regression line is: $(a_2/a_1)_{\{e({\bf C1})\}}= 0.0643 M_{i} + 1.5797$. 
The crosses mark the observed spectra plotted in Figure~\ref{fig:extraDC1_a2_a1_realspec.2}a.
\label{fig:extraDC1_a2_a1_redshift}}
\end{figure}

Luminosity effects on broad emission lines can also be probed in a
similar way to the cross-redshift projection. One prominent
luminosity effect is found by projecting $spec({\bf D1})$ onto $\{e({\bf C1})\}$. These samples
have the same redshift range but different luminosities (for {\bf D1}, $M_{i}=(-24, -22)$
and for {\bf C1}, $M_{i}=(-26, -24)$).
Figure~\ref{fig:extraDC1_a2_a1_redshift} shows the eigencoefficient $(a_2/a_1)_{\{e({\bf C1})\}}$
as a function of absolute luminosity, with redshifts fixed at $z=0.4 \pm 0.02$ (235 quasars). The ratio of the
first 2 eigencoefficients decreases with increasing quasar luminosity. The regression line (with
outliers of $a_2/a_1 > 1$ removed from the calculation) is 
\begin{equation}
(a_2/a_1)_{\{e({\bf C1})\}}= 0.0643 M_{i} + 1.5797 \ ,
\end{equation}

\noindent 
with a correlation coefficient of 0.2305 with an extremely significant two-tailed P-value of 0.0003.

\begin{figure}
\plotone{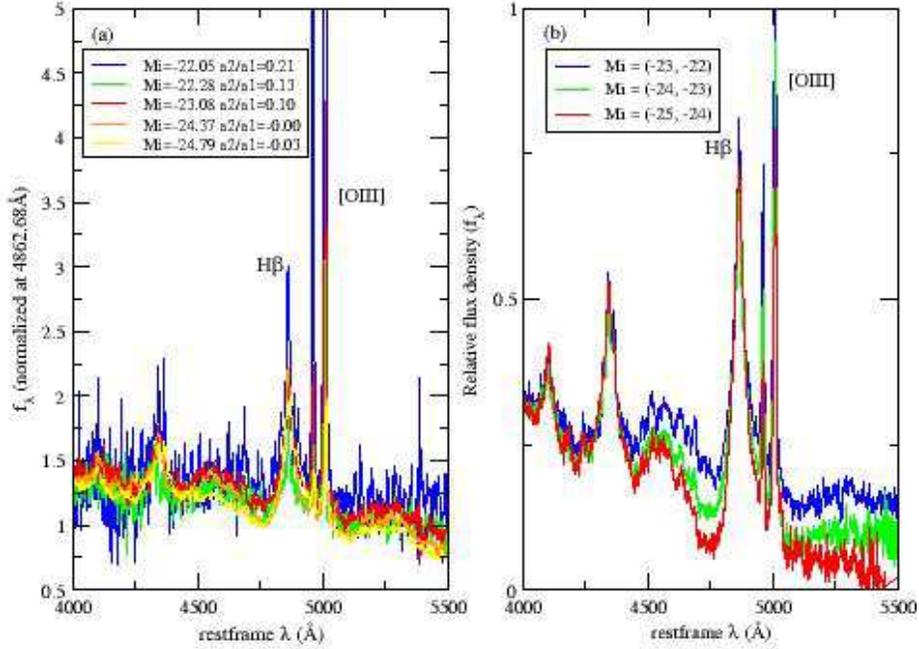}
\figcaption[fig19.eps]{(a) The $H{\beta}$ region of spectra located along the regression line 
for $(a_2/a_1)_{\{e({\bf C1})\}}$ versus $M_{i}$ (marked with crosses in Figure~\ref{fig:extraDC1_a2_a1_redshift}). 
The spectra are normalized by the continuum flux density at $\lambda=4862.68$~\AA \, and 
smoothed with a FWHM $=3$~\AA \, Gaussian smoothing function. 
For decreasing $a_2/a_1$ ratio, i.e., for brighter quasars, the emission line equivalent
widths typically decrease. (The QSOs are, starting from the brightest,
\mbox{SDSS J013418.19+001536.6}, \mbox{SDSS J010342.73+002537.2}, \mbox{SDSS J011310.38-003133.1}, 
\mbox{SDSS J093409.17+023237.0}, \mbox{SDSS J092011.60+571718.2}.) 
(b) The geometric composite spectra in different absolute luminosity bins
 with $M_{i}$ ranges from -22 to -25 for the objects in Figure~\ref{fig:extraDC1_a2_a1_redshift}.
(Lowered Resolution) \label{fig:extraDC1_a2_a1_realspec.2}} 
\end{figure}

Along this luminosity trend, the equivalent widths of emission lines such as H$\beta$
and $[$\ion{O}{3}$]$ lines are found to decrease typically, 
as a function of increasing absolute magnitude $M_{i}$ (as shown in the spectra in 
Figure~\ref{fig:extraDC1_a2_a1_realspec.2}a). This is the Baldwin (1977) effect. We note that
the host-galaxy may come into play in this case (at low redshifts and low luminosities). 
The geometric composite spectra of different
luminosities within the range from $-22$ to $-25$ are shown in Figure~\ref{fig:extraDC1_a2_a1_realspec.2}b,
in which a spectral index of $\alpha_{\nu} = -0.5$ for the continua is assumed. The Baldwin effect for the
emission lines is also present.

In the highest redshift bins, the Baldwin effect can be found in the first and 
the second eigenspectra. Figure~\ref{fig:ZBIN4_eigenspec} shows that the addition (with positive eigencoefficients) of 
the first two eigenspectra {\it enhances} the flux density around 1450~\AA \, and reduces the equivalent width 
of \ion{C}{4}. Ly$\alpha$ and other major BELs are also shown to be
anti-correlated with the continuum flux. Hence, the Baldwin effect is not limited
to the \ion{C}{4} emission line, and is also observed in many broad emission lines (see, for example, a summary in 
Sulentic {et~al.} 2000). The linear-combination of the first and third modes in this redshift range also shows a
similar modulation between the flux density around 1450~\AA \, and the line equivalent width. This
effect is, however, not general for all luminosities, with the third eigenspectrum in 
{\bf C4} showing only a small value in the 1450~\AA \, flux density.

\begin{figure}
\epsscale{0.6}
\plotone{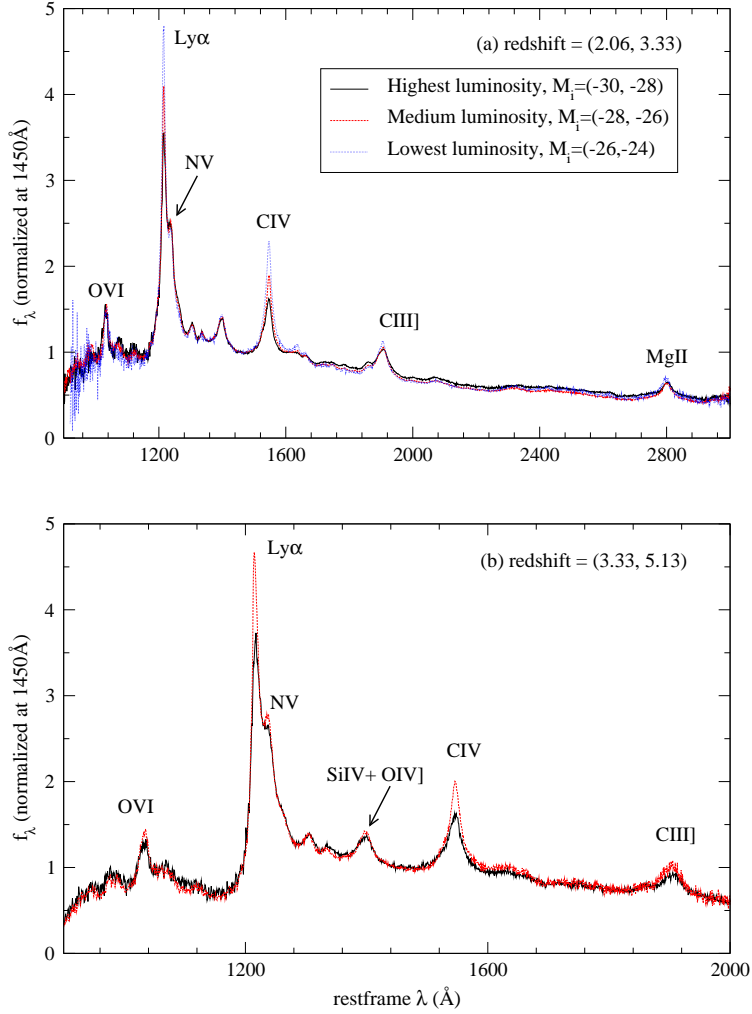}
\figcaption[fig20.eps]{Baldwin effect of the broad emission lines \ion{O}{6} (only in the highest redshifts), 
Ly$\alpha$, the blended \ion{Si}{4} + \ion{O}{4}$]$, \ion{C}{4}, \ion{He}{2}, \ion{C}{3}$]$ and \ion{Mg}{2} 
are shown in the redshift ranges (a) $2.06-3.33$ and (b) $3.33-5.13$. In both cases, the 1st eigenspectra in 
different luminosity ranges are shown, and are normalized at 1450~\AA. 
For \ion{N}{5}, the effect appears to be redshift dependent 
(see \S~\ref{section:baldwin} for discussions). \label{fig:baldwin_CIV}}
\end{figure}

The Baldwin effect can also be seen by comparing the first eigenspectra constructed
for different luminosity bins. Figure~\ref{fig:baldwin_CIV} shows the first eigenspectra derived in different 
luminosities in the second highest redshift bin (i.e., the $(M_{i}, z)$-bins {\bf A4}, {\bf B4} and {\bf C4}, 
with $2.06 < z < 3.33$) and the highest one ({\bf A5} and {\bf B5,} with $3.33 < z < 5.13$). The 
eigenspectra are normalized to unity at $1450$~\AA. The continua for wavelengths
approximately greater than 1700~\AA \, in Figure~\ref{fig:baldwin_CIV}a are not perfectly normalized
 (which is difficult to define in the first place),  but a more careful normalization
would only lead to an increase in the degree of the Baldwin effect in the emission lines \ion{C}{3}$]$ and \ion{Mg}{2}.
The Ly$\alpha$ and \ion{C}{4} lines demonstrate the most profound Baldwin effect. 
Other broad emission lines such as \ion{He}{2}$\lambda$1640, 
\ion{C}{3}] and \ion{Mg}{2} also exhibit this effect. For the controversial line \ion{N}{5}, 
an ``anti-Baldwin'' correlation is found at redshifts $2.06-3.33$, such that 
flux densities are smaller for lower-luminosity quasars. At the highest redshifts in 
this study ($z=3.33-5.13$, Figure~\ref{fig:baldwin_CIV}b), however, a normal Baldwin effect of 
\ion{N}{5} is found. The redshift dependency in the Baldwin effect for \ion{N}{5} may explain
the contradictory results found in previous studies (a detection of Baldwin effect of \ion{N}{5} in Tytler \& Fan 1992;
and non-detections in Steidel \& Sargent 1991; Osmer {et~al.} 1994; and Laor {et~al.} 1995).
While most studies have shown little evidence of
the Baldwin effect in the blended emission lines \ion{Si}{4}+\ion{O}{4}$]$, our results support the existence
of an effect (though at a much weaker level than that of Ly$\alpha$ and \ion{C}{4}).
This is in agreement with two previous works (Laor {et~al.} (1995) which used 14 HST 
QSOs, and Green, Forster \& Kuraszkiewicz (2001) which used about 400 QSOs from the LBQS). 
In the optical region, at least \ion{He}{2}$\lambda$4687 was reported to show the Baldwin 
effect \cite{Heckman80, Boroson92, Zheng93}.

To further verify that the luminosity dependency of the eigencoefficients implies a Baldwin effect, 
we also study the eigencoefficients corresponding to the Baldwin effect seen in Figure~\ref{fig:baldwin_CIV}. 
We find that when $spec({\bf C4})$ are 
projected onto $\{e({\bf B4})\}$ the luminosity dependency is also seen in the eigencoefficients, with
$(a_2/a_1)_{\{e({\bf B4})\}}= 0.0327 M_{i} + 0.8794$ ($r = 0.1150$, and an insignificant two-tailed P-value of 0.14)
and $(a_3/a_1)_{\{e({\bf B4})\}}= -0.0616 M_{i} - 1.6948$ ($r = 0.2177$, and
a very significant two-tailed P-value of 0.0043), 
both for objects with redshifts within $2.7 \pm 0.1$
(161 objects in the case of $a_2/a_1$ and 166 in that of $a_3/a_1$).

\section{A Spectral Sequence along Eigencoefficients $(a_1, a_2)$ in $(M_{i}, z)$-bin} \label{section:class}

\begin{figure}
\epsscale{0.8}
\plotone{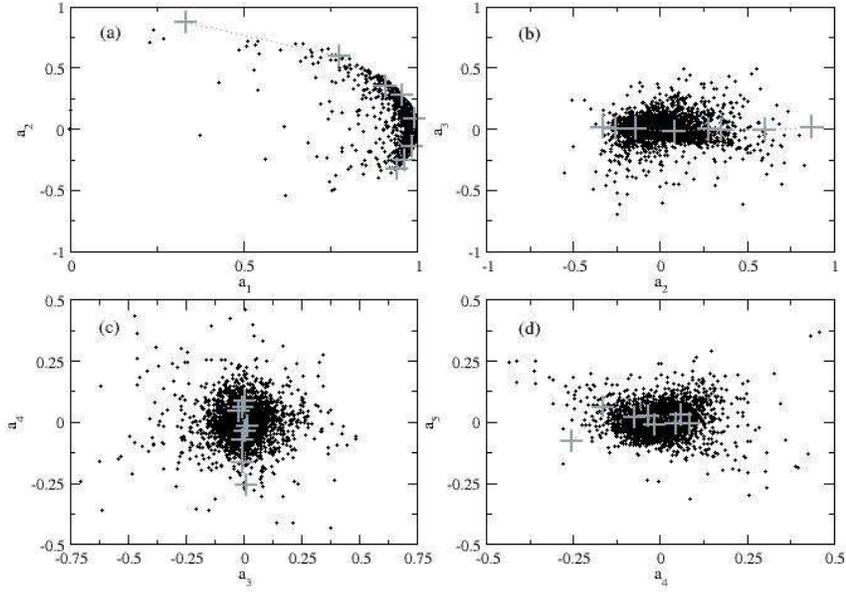}
\figcaption[fig21.eps]{The distributions of the first 5 eigencoefficients of the $(M_{i}, z)$-bin {\bf B3},
in (a) $a_2$ versus $a_1$ (b) $a_3$ versus $a_2$ (c) $a_4$ versus $a_3$ and (d) $a_5$ versus $a_4$. 
The crosses mark the observed QSO spectra illustrated in Figure~\ref{fig:BZBIN3_pickspec_a1_a2}. (Lowered Resolution)
\label{fig:BZBIN3_a1_a5}} 
\end{figure}

Figure~\ref{fig:BZBIN3_a1_a5} shows plots of the first five eigencoefficients of the $(M_{i}, z)$-bin 
{\bf B3}, where the properties are typical for all $(M_{i},z)$-bins. The eigencoefficients
are normalized as: $\sum_{m=1}^{50} a_{m}^{2} = 1$.
The plot of $a_2$ versus $a_1$ shows a continuous progression in the ratio of these coefficients which 
is similar to that found in the KL spectral classification
of galaxies \cite{Con95}, in which the points fall onto a major ``sequence'' of increasing spectral slopes. 
As higher orders are considered, for example $a_5$ vs $a_4$ (Figure~\ref{fig:BZBIN3_a1_a5}d), 
no significant correlations are observed.

\begin{figure}
\epsscale{0.68}
\plotone{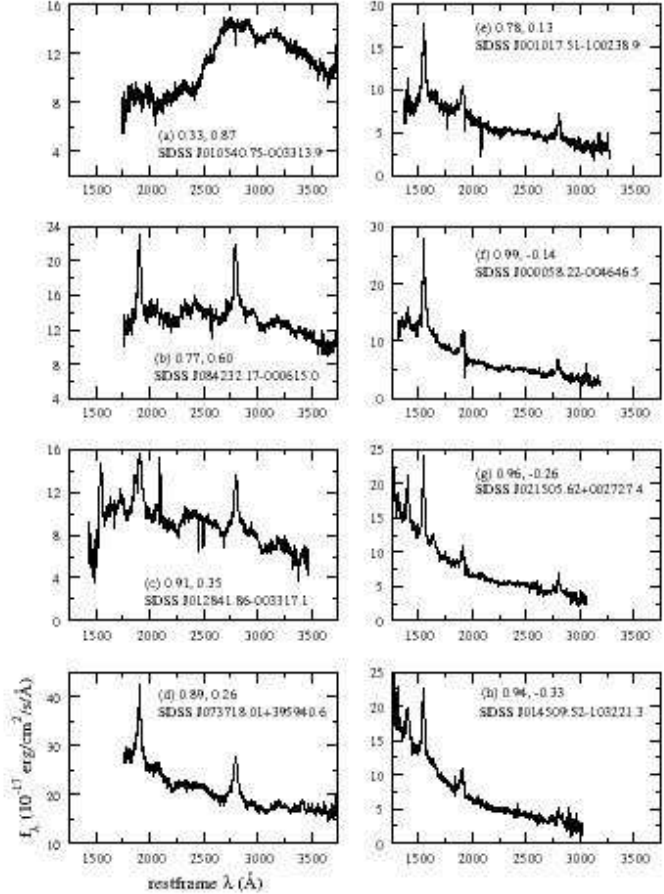}
\figcaption[fig22.eps]{The observed quasar spectra picked along a sequence 
formed by the first two eigencoefficients $a_1$ and $a_2$ valued at the two numbers
in each figure (see Figure~\ref{fig:BZBIN3_a1_a5} for actual locations of $a_1$ to $a_5$). 
Proceeding along the sequence, the spectral slopes of quasars
progressively vary from redder to bluer. The spectra are smoothed
by a FWHM $=3$~\AA \, Gaussian smoothing function for easier visualization. The $(M_{i}, z)$-bin under consideration
is {\bf B3}. (Lowered Resolution) \label{fig:BZBIN3_pickspec_a1_a2}} 
\end{figure}

Observed quasar spectra are inspected along this trend of $a_2$ versus
$a_1$ (Figure~\ref{fig:BZBIN3_pickspec_a1_a2}).  The top of each
sub-figure shows the values of $(a_1, a_2)$. Along the sequence with
decreasing $a_2$ values, the quasar continua are progressively
bluer. The relatively red continua in
Figures~\ref{fig:BZBIN3_pickspec_a1_a2}a to
\ref{fig:BZBIN3_pickspec_a1_a2}c  may be due
intrinsic dust obscuration \cite{Hall02}.
The quasar in Figure~\ref{fig:BZBIN3_pickspec_a1_a2}c 
is probably a high-ionization BALQSO (HiBAL) according to the supplementary SDSS EDR BAL
quasar catalog \cite{Reichard03}. We do, however, emphasize that the appearance of
this BALQSO (or any BALQSO in general) in this particular sequence of quasar
in the $a_2$ versus $a_1$ plane does not imply two modes
are enough to achieve an accurate classification for a general BALQSO (for the reasons described in 
\S~\ref{section:BAL}). The steepness of the spectral slope of this particular BALQSO is the major reason 
which causes such values of $a_1$ and $a_2$ eigencoefficients.

On the variations of the emission lines along these major $(M_{i},z)$
sequences, we can appreciate some of the difficulties in obtaining a
{\it simple} classification concerning {\it all} emission lines by inspecting
the examples listed in Table~\ref{tab:FWHMzbin}. The addition of the 2nd eigenspectrum 
to the 1st, weighted with (signed) medians of the eigencoefficients for all objects in a given sample,
broadens some emission lines while making others
narrower; a similar effect is seen for the addition of the 3rd eigenspectrum to the 1st, but
in two {\it different} sets of lines. This shows the large intrinsic variations in the
emission line-widths of the QSOs.

\section{Local Eigenspectra and Correlations among Emission Lines} \label{section:linecorr}

One of the utilities of the KL transform is to study the linear
correlations among the input parameters, in this case, the pixelized
flux densities in a spectrum.  Due to possible uncertainties
in any continuum fitting procedure in quasar spectra and the fact that no
quasar spectrum in our sample completely covers the rest wavelength range
$900-8000$~\AA, correlations among the broad emission lines are first
determined locally around the lines of interest by studying the first
two eigenspectra in a smaller restricted wavelength range using the wavelength-selected QSO spectra. 
This process is then repeated from 900~\AA \, to 8000~\AA.
 Each local wavelength region is chosen to be  $\approx 500-1800$~\AA \, wide in
the restframe. Empirically, we find that at these spectral
widths the correlations among broad emission lines can be isolated in the
first two eigenspectra without interference by the continuum information
(except in the vicinity of \ion{Mg}{2} doublet, for which the adjacent strong
emission lines are located well beyond the \ion{Fe}{2} (UV) region, which can be as
broad as $\lambda_{rest} \approx$ $2000-4000$~\AA), {\it in contrast} to the property of the $(M_{i}, z)$-bins 
in which the 2nd eigenspectra generally describe the variations in the spectral slopes.

The actual procedures to determine the correlations among the strengths of the major
emission lines are as follows: $(i)$ in each bin, the eigencoefficients of all objects
are computed, and the distribution of the first two eigencoefficients, $a_2$ versus $a_1$,
are divided into several ($\approx 10$) sections within $\pm 1 \sigma$ of the $a_2$ distribution. In  
each section the mean eigencoefficients, $<a_1>$ and $<a_2>$, are  calculated (discarding outliers $a_1 < 0$). 
$(ii)$ Along this trend of mean eigencoefficients, synthetic 
spectra are constructed by the linear-combination of the first two eigenspectra
using the weights defined by the mean eigencoefficients. $(iii)$ The equivalent widths of emission lines in
the synthetic spectra are calculated along the trend of mean eigencoefficients, so that
the correlations among the strengths of the broad emission lines can be deduced. Linear regression 
and linear correlation coefficients are calculated from the EW-sequence of a particular emission line relative to
 that of another line, which is fixed to be the emission line with the shortest 
wavelength of each local bin. The equivalent widths
are calculated by direct summation over the continuum-normalized flux densities within
appropriate wavelength windows. From such procedures, the correlations found are   
ensemble-averaged properties of redshifts and luminosities over the corresponding range, 
and are physical. Table~\ref{tab:linedata} shows the rest-wavelength bounds, the redshift range,
the number of quasar spectra in each bin, and regression and correlation coefficients for each major emission line. 
The range of the possible restframe equivalent widths (EW$_{rest}$) along $(a_1, a_2)$ is listed in decreasing
$a_2$ values. Since the 
redshifts are chosen such that each quasar spectrum has a full coverage in the corresponding 
 wavelength region, the gap-correcting procedure is implemented to correct only for 
skylines and bad pixels. 

The EW$_{rest}$ of the emission lines vary at different magnitudes along the $(a_1, a_2)$ sequence; some
change by nearly a factor of two (e.g., Ly$\alpha$, \ion{C}{4}), while some show smaller changes
(e.g., \ion{Si}{4}+\ion{O}{4}$]$, \ion{C}{3}$\lambda$1906). Within a single
local bin, the rest equivalent widths of some emission lines increase while others decrease along
the trend $(a_1, a_2)$ with decreasing $a_2$ values. These results are the testimonies to the fact
that quasar emission lines are diverse in their properties.

We also note that some pairs of emission lines change their correlations as a function
of redshift (i.e., different local bins). 
For example, \ion{Mg}{2} is correlated with \ion{O}{3}+\ion{Fe}{2}(Opt82) in the 
local bin of  $z = 1.1 - 1.87$ but anti-correlated in that of $z = 0.46 - 1.16$. Another example is
the $[$\ion{S}{2}$]$$\lambda6718$ and $[$\ion{S}{2}$]$$\lambda6733$ pair. Hence if  correlations are interpreted
 between the emission lines from one local bin with those from an adjacent bin, 
caution has to be exercised. The uncertainty in the continuum estimation
(e.g., the iron contamination in the continuum in the vicinity of \ion{Mg}{2})
prevents us from drawing an exact physical interpretation of this phenomenon.

\subsection{Francis's PCs and Boroson \& Green's ``Eigenvector-1'' in SDSS QSOs}

\begin{figure}
\epsscale{0.65}
\plotone{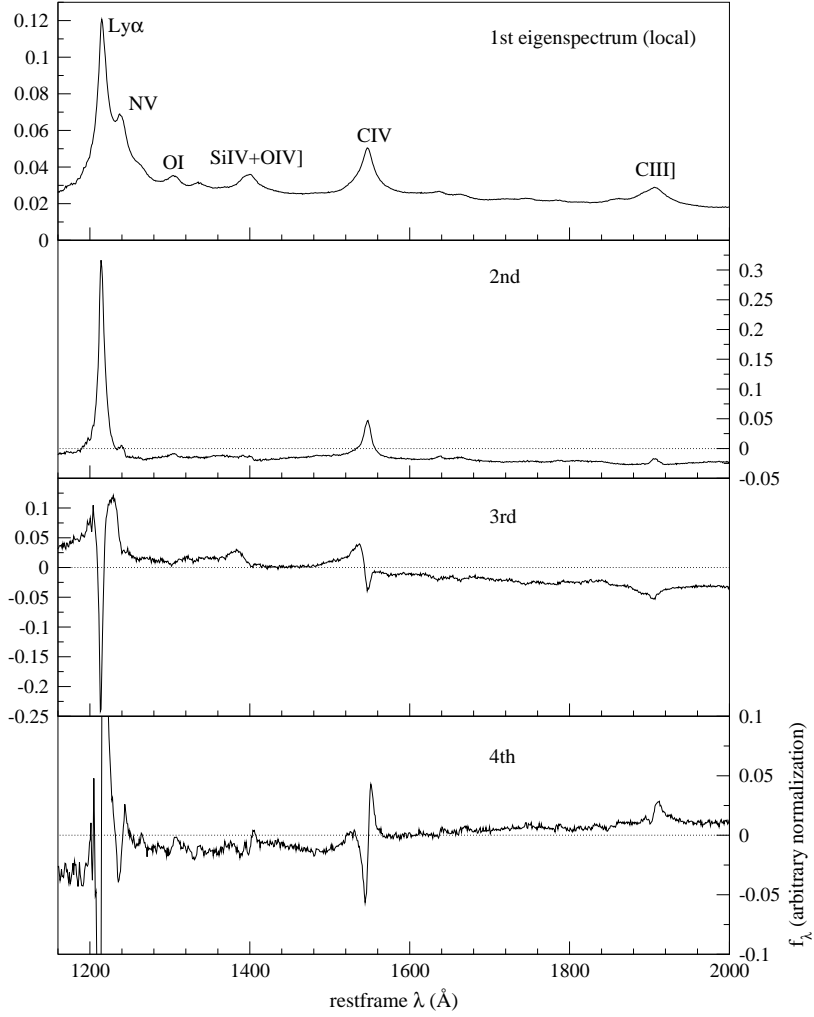}
\figcaption[fig23.eps]{The first 4 orders of locally constructed eigenspectra in the restricted 
wavelength region $1150 -2000$~\AA, in which the quasar spectra are chosen to have full wavelength coverage. In this 
rather narrow rest-wavelength coverage, the emission lines in the 2nd eigenspectrum show the low-velocity core 
components (i.e., relatively narrower than the first eigenspectrum). The
correlations among the relevant board emission lines are probed in the 2nd eigenspectrum. The results agree 
very well with that by Francis {et~al.} (1992), in, for example, the correlation between  Ly$\alpha$ and 
\ion{C}{4}; and the information about the variations of continuum slopes showing up in the 3rd eigenspectrum. 
\label{fig:Lyalpha_eigenspec}}
\end{figure}

\begin{figure}
\epsscale{0.65}
\plotone{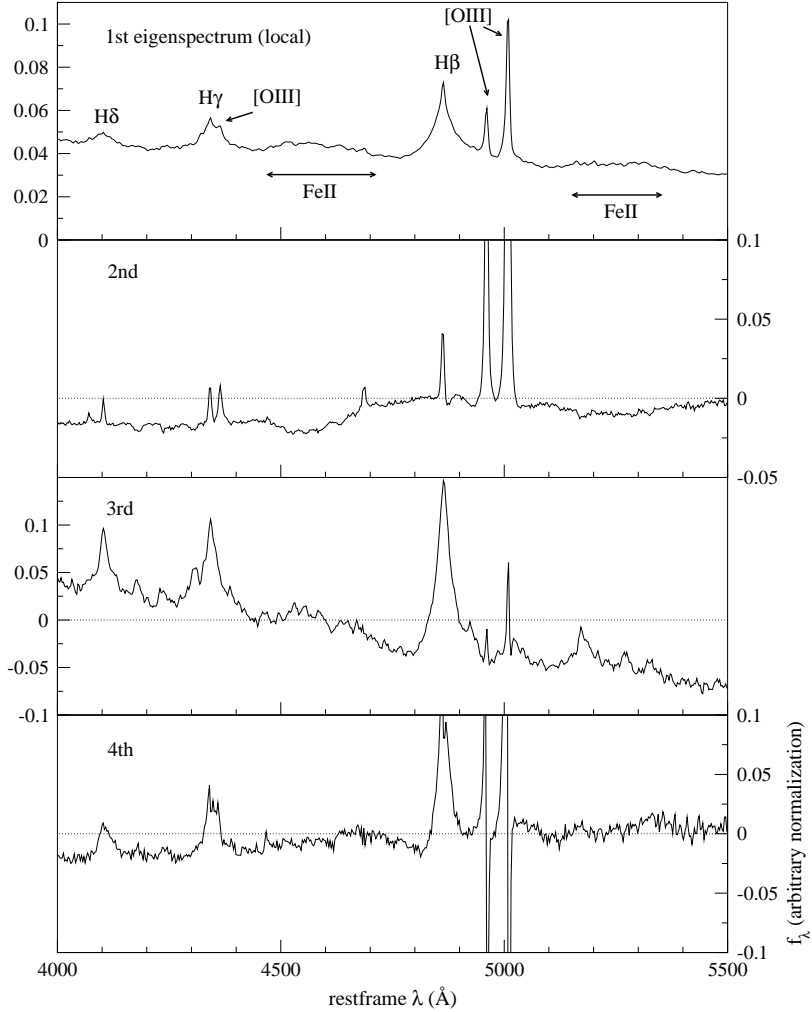}
\figcaption[fig24.eps]{The first 4 orders of locally constructed eigenspectra in the restricted 
wavelength region $(4000, 5500)$~\AA, in the vicinity of H${\beta}$. The well-known ``Eigenvector-1'' 
(Boroson and Green, 1992), which essentially is the anti-correlation between [\ion{O}{3}] and \ion{Fe}{2} (optical),
 are clearly shown in the 2nd eigenspectrum in our work, which is enlarged for easier visualization. 
The blended H$\gamma$ and [\ion{O}{3}] in the 1st eigenspectrum is cleanly split in the 
2nd eigenspectrum, which is a nice example showing that the locally constructed 2nd eigenspectrum 
comprises mainly the line-core components (see also Figure~\ref{fig:linecore}). 
The 3rd eigenspectrum shows prominent contributions from Balmer lines as well as the continuum slope. 
\label{fig:Hbeta_eigenspec}} 
\end{figure}

Two examples of the locally-constructed eigenspectra 
 are shown in Figures~\ref{fig:Lyalpha_eigenspec} and \ref{fig:Hbeta_eigenspec}.
In Figure~\ref{fig:Lyalpha_eigenspec}, the eigenspectra are constructed using wavelength-selected QSO spectra in the 
rest-wavelengths $1150-2000$~\AA \, (with $2.3<z<3.6$), so that both Ly$\alpha$ and \ion{C}{4} are covered. 
Excellent agreement is shown between our eigenspectra and those selected from 
the Large Bright Quasar Survey in the $1.8<z<2.7$ range \cite{Francis92}. The second eigenspectrum
 (corresponding to the first principal component in Francis {et~al.}) shows the line-core
components of emission lines. In contrast, the 3rd mode (corresponding to their 2nd principal component) 
 shows the
continuum slope, with the node located at around 1450~\AA. Besides, the addition (with positive eigencoefficient) of
the 3rd eigenspectrum to the 1st one enhances the fluxes at shorter wavelengths
while {\it increases} the \ion{C}{4} blueshift. This supports the finding of a previous
study \cite{Richards02b} that \ion{C}{4} blueshift is greater in bluer SDSS QSOs.

At longer wavelengths, the SDSS quasars with redshifts $0.08-0.67$ show the anti-correlation between 
\ion{Fe}{2} (optical) and [\ion{O}{3}] (Figure~\ref{fig:Hbeta_eigenspec}), in agreement with the Eigenvector-1 
(Boroson \& Green 1992). The first two eigenspectra in Figure~\ref{fig:Hbeta_eigenspec} demonstrate
that both the H$\beta$ and the nearby [\ion{O}{3}] forbidden lines are anti-correlated with the 
\ion{Fe}{2} (optical) emission lines, which are the  
blended lines blueward of H$\beta$ and redward of [\ion{O}{3}]. 
In the 3rd local eigenspectrum, the Balmer emission lines are prominent, which was noted
previously in the PCA work by Shang {et~al.} (2003). In addition, we find a correlation between
the continuum and the Balmer lines in this local 3rd eigenspectrum, 
so that their strengths are stronger in bluer quasars.
 
To date, it is generally believed that the anti-correlation between \ion{Fe}{2} (optical) and [\ion{O}{3}] 
is not driven by the observed orientation of the quasar. One of the arguments by Boroson \& Green was
that the [\ion{O}{3}]$\lambda$5008 luminosity is an isotropic property. Subsequent studies of radio-loud AGNs have 
put doubt on the isotropy of the [\ion{O}{3}] emissions. Recent work by Kuraszkiewicz et~al. (2000), however, showed a 
significant correlation between Eigenvector-1 and the evidently orientation-independent [\ion{O}{2}] emission
in a radio-quiet subset of the optically selected Palomar BQS sample, 
which implies that external orientation probably does not drive the Eigenvector-1. An interesting future
project to address this problem is to relate the quasar eigenspectra in the SDSS to their
 radio properties.

\subsection{Weight of Line-Core}

\begin{figure}
\epsscale{0.8}
\plotone{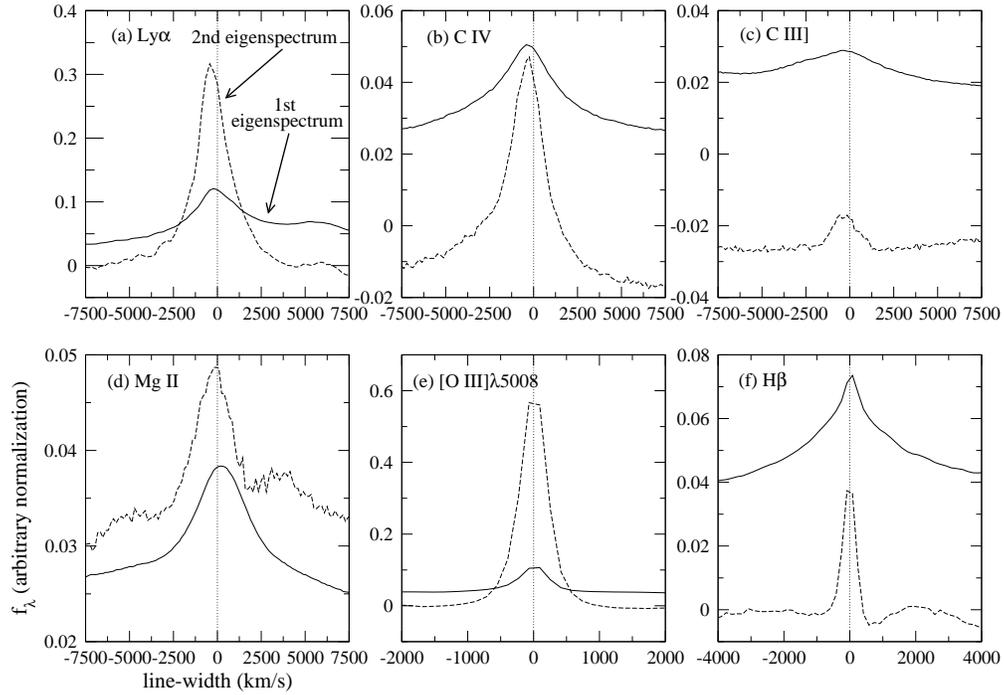}
\figcaption[fig25.eps]{The first (solid line) and the second (dashed line) locally constructed eigenspectra
in the regions around (a) Ly$\alpha$ (b) \ion{C}{4} (c) \ion{C}{3}$]$ (d) \ion{Mg}{2} 
(e) $[$\ion{O}{3}$]$$\lambda$5008 and (f) H$\beta$. The generally narrower second eigenspectrum 
(except for \ion{Mg}{2}, in which the conclusion is complicated by the surrounding \ion{Fe}{2} lines) 
compared with the first one suggests that it mainly carries the line-core (i.e., low radial velocity) 
information to varies levels. \label{fig:linecore}} 
\end{figure}

Enlargements of the first two locally constructed eigenspectra focusing
on  major broad emission lines are illustrated in
Figure~\ref{fig:linecore}. Except for the almost perfectly symmetric and
zero velocity of the line centers of the 1st and 2nd eigenspectra
exhibited by [\ion{O}{3}]$\lambda$5008, most broad emission lines do show
asymmetric and/or blueshifted profiles. These demonstrate the variation of
broad line profiles of quasars and the generally blueshifted broad
emission lines relative to the forbidden narrow emission lines.  The
forbidden lines in the narrow line regions of a QSO are always adopted in calculating the
systemic host-galaxy redshift, so the clouds associated with
blueshifted BELs probably have additional velocities relative to the host. This
line-shift behavior was found in many other studies (see references in Vanden~Berk et~al. 2001).
The behavior of the \ion{C}{4} shift led Richards et~al. (2002b) to suggest that orientation
(whether external or internal) may be the cause of the effect.

It is also obvious from Figure~\ref{fig:linecore} that the 2nd eigenspectra are generally narrower (except for
\ion{Mg}{2}, in which the conclusion is complicated by the presence of the surrounding \ion{Fe}{2} 
lines) than their 1st eigenspectra counterparts.
The line-widths of the sample-averaged KL-reconstructed spectra using only the first eigenspectrum or
the first two eigenspectra are listed in Table~\ref{tab:FWHMlocal}. The addition of the first two modes, 
weighted by the medians of the eigencoefficients, causes the widths of 76~\% 
of the emission lines (with FWHM $> 1000$~km~s$^{-1}$) to be narrower than those reconstructed from the 
first mode only. Hence, most broad emission lines
can be mathematically decomposed into broad, high-velocity components and narrow, low-velocity components.
Appearing in the second local eigenspectra, the line-widths are thus the most important 
variations of the quasar broad emission lines. The line-core components were reported by Francis {et~al.} (1992) for 
 \ion{C}{4} and Ly$\alpha$; and Shang {et~al.} (2003) for some major broad emission lines.
One nice illustration of the line-core component of the 2nd mode is  
the splitting of H$\gamma$ and its adjacent [\ion{O}{3}] in Figure~\ref{fig:Hbeta_eigenspec},
for they are blended in the 1st mode.

Similar properties may be expected in the 2nd $(M_{i}, z)$-binned eigenspectra.
Table~\ref{tab:FWHMzbin} lists the average FWHM of different linear combinations using the 
first 3 eigenspectra in constructing some major broad emission lines. 
Comparatively, for most emission lines the second $(M_{i}, z)$-binned eigenspectra do not 
show as narrow line components as the second eigenspectra, in which the widths 
of 61~\% of the emission lines with FWHM $> 1000$~km~s$^{-1}$
 become narrower by adding the 2nd eigenspectrum to the 1st one.
This effect is mainly due to the difference in the numbers of quasars, and  more importantly,
the inclusion of a wider spectral region causes the ordering of the weights of different physical properties to 
re-arrange. In this case, the spectral slope variations are more important than those of the line-cores.
While the 3rd $(M_{i},z)$-binned eigenspectra (weighted by medians of the eigencoefficients of the sample) 
also do not represent prominent changes in the emission line-cores,
except for Ly$\alpha$ and \ion{C}{4} (the FWHM of \ion{C}{4} appears to be larger
because the line-core 3rd mode is pointing downward in {\bf ZBIN~4}), on average the 
quasar populations with {\it negative} 3rd eigencoefficients do show narrower widths for 
77~\% of the emission lines. Similarly, the 2nd global eigenspectrum does not carry dominant emission 
line-core components, which are found to be represented more prominently by the 3rd mode
(Table~\ref{tab:FWHMglobal}).

\subsection{FWHM-EW Anti-Correlation in BELs: Classification?} \label{section:FWHM_EWclass}

The narrower emission features in the 2nd local eigenspectrum compared with the 1st 
one, and the fact that almost every broad emission line is pointing towards  positive flux values
in both of these two modes, imply that there is an anti-correlation between FWHMs and the 
equivalent widths of broad emission lines.
In fact, as suggested by Francis {et~al.} (1992), this
may form a basis for the classification of quasar spectra in $\lambda_{rest}=1150-2000$~\AA, by arranging them 
accordingly into a sequence varying from narrow, large-equivalent-width  to broad, 
low-equivalent-width emission lines. From the locally constructed eigenspectra,
such an anti-correlation is not generally true for every broad emission line as we find that
there exists at least one exception: a positive correlation between the FWHM and the EW of \ion{Mg}{2} 
in the local bin of the redshift range $0.46-1.16$. An assumption in these measurements is that 
the continuum underneath can be approximated by a linear-interpolation across the window $2686 - 2913$~\AA.
One complication, however, is 
the contamination due to the many \ion{Fe}{2} emission lines in the vicinity of  \ion{Mg}{2}, so the true continuum 
may be obscured. The positive FWHM-EW correlations appear to exist in some other weaker
emission lines as well, but the weak strengths of those lines do not permit us to draw definitive conclusions 
under the current spectral resolution. In conclusion, the FWHM-EW relation can help us to classify
most broad emission lines individually, but 
{\it this relation cannot be used in a general sense, nor does it represent the most important sample variation},
if the surrounding continua
are included to the extent of the rest-wavelength ranges of the $(M_{i},z)$-binned spectra. Nonetheless,
most broad emission lines can be viewed mathematically as the combinations of
broad and narrower components. A future study will focus on 
finding the best physical parameters for classifying the spectra in the wide spectral region, which will be
the subject of a second paper. One possible approach is to study the distributions of the eigencoefficients 
and their relations with other spectral properties (e.g., Francis et~al. 1992; Boroson \& Green 1992).

\subsection{Local Spectral Properties in the $(M_{i}, z)$-Eigenspectra} \label{section:localinMizbin}

The shapes of the continua and the correlations among the broad
emission lines of the second  locally constructed eigenspectra
are all identified in either the 3rd or the 4th $(M_{i}, z)$-binned eigenspectra.
We do expect, and it is indeed found to be true, that the local properties of the spectra
can be found in the latter, though the ordering may be different. The identifications 
are marked in Figures~\ref{fig:ZBIN1_eigenspec} $-$ \ref{fig:ZBIN5_eigenspec}
by the redshift ranges of the local eigenspectra,
with reference to the luminosity averaged {\bf ZBIN} eigenspectra. 
The correlations of broad emission lines
are generally found in higher-order $(M_{i}, z)$-binned eigenspectra compared with the orders
representing the spectral slopes. 

\section{Summary and Future work} \label{section:conclusion}

We perform KL transforms and gap-corrections on 16,707 SDSS  quasar spectra. 
In rest-wavelengths $900 - 8000$~\AA, the 1st eigenspectrum (i.e., the mean spectrum) shows agreement
with the SDSS composite quasar spectrum \cite{VandenBerk01}, with an abrupt change in the spectral slope 
around 4000~\AA. The 2nd eigenspectrum carries the host-galaxy contributions to the quasar spectra, 
hence the removal of this mode can probably prevent the obscuration of the real physics of galactic 
nuclei by the stellar components. Whether this eigenspectrum is the only one containing galaxy 
information requires further study. The 3rd eigenspectrum  shows the modulation between the UV and the 
optical spectral slope, in agreement with the 2nd principal component of Shang {et~al.} (2003). 
The 4th eigenspectrum  shows the correlations between Balmer emission lines. 

Locally around various broad emission lines, the eigenspectra from the wavelength-selected quasars 
qualitatively agree with those from 
the Large Bright Quasar Survey, the properties in the Eigenvector-1 \cite{Boroson92},
and the anti-correlations between the FWHMs and the equivalent widths of Ly$\alpha$ and \ion{C}{4} \cite{Francis92}.
The anti-correlation between the FWHM and the equivalent width is found in most 
broad emission lines with few exceptions (e.g., \ion{Mg}{2} is discrepant).

From the commonality analysis of the subspaces spanned by the eigenspectra in different
redshifts and luminosities, the spectral classification of quasars is shown 
to be redshift and luminosity dependent. 
Therefore, we can either use of order 10 
$(M_{i}, z)$-binned eigenspectra, or of order $100$ global eigenspectra to represent most
 (on average 95~\%) quasars in the sample. 
We find that the first two modes can describe the spectral slopes of the quasars in  
all $(M_{i}, z)$-bins under study, which is the most significant sample variance of the current QSO catalog.
The simplest classification scheme can be achieved  based on the first two eigencoefficients, so that
a physical sequence can be formed upon the linear-combinations of the first two eigenspectra.
The diversity in quasar spectral properties, and the inevitable different restframe wavelength 
coverages due to the nature of the survey, increase the sparseness of the data. Hence, higher-order modes
enter into the construction of the broad emission lines 
with the eigenspectra, in contrast to the galaxy spectral classification, in which
most emission lines vary monotonically with the spectral slope \cite{Con95}. This result is
also a manifestation of the high uniformity of galaxy spectra compared with quasar spectra.

We find that BAL features do not only appear in one particular order of eigenspectrum
but span a number of orders, mainly higher-orders. 
This may indicate substantial challenges to the classification of BAL quasars
by the current sets of eigenspectra in terms of arriving at a compact description. 
A separate KL-analysis of the BAL quasars is desirable
for studying the classification problem. Nonetheless, the appropriate truncation of the number of eigenspectra 
in reconstructing a quasar spectrum can in principle lead to an un-absorbed continuum.

We find evolution of the small bump by the cross-redshift KL transforms,
in agreement with the quasars from the Large Bright Quasar Survey
\cite{Green01} and in other independent work \cite{Kuhn01}.  The
Baldwin effect is detected in the cross-luminosity KL transforms, as well as from the mean QSO spectra derived 
for different luminosities.  One implication of these redshift and
luminosity effects is that they have to be accounted for in the spectral classification of quasars,
consistent with our finding from the commonality analysis.

The high quality of the data allows us to obtain quasar eigenspectra
which are generic enough to study spectral properties. Despite the
presence of diverse quasar properties such as different continuum slopes and
shapes, and various emission line features known for several decades, 
our analysis shows that there are unambiguous correlations among various broad emission 
lines and with continua in different windows.

A second paper is being prepared to address the classifications of the
DR1 quasars in greater detail. One interesting direction is to relate
the current eigenspectra approach to the radio properties of the
quasars, so that further discriminations of intrinsic and extrinsic
properties can be achieved, for example, the orientation effects on the
observed spectra (e.g., Richards et~al. 2002b). Another application currently being addressed is the
removal of host-galaxy components from the SDSS quasar spectra.  In
addition, the cross-projections can also be applied to study future larger
samples of quasars (e.g., $\approx$ 100,000 at the completion of the SDSS) for
possibly new evolution and luminosity effects. 

\acknowledgements

We thank David~Turnshek for the discussion of the Baldwin effect. We thank
Ravi~Sheth for various comments and discussions. 
We thank the referee Zhaohui~Shang for the helpful comments.
CWY and AJC acknowledge partial support from
an NSF~CAREER~award~AST99~84924 and a NASA~LTSA~NAG58546. AJC
acknowledges support from an NSF ITR awards AST-0312498 and ACI-0121671.

Funding for the creation and distribution of the SDSS Archive has been provided by the
Alfred P. Sloan Foundation, the Participating Institutions, the National Aeronautics and
Space Administration, the National Science Foundation, the U.S. Department of Energy, the
Japanese Monbukagakusho, and the Max Planck Society. The SDSS Web site is
http://www.sdss.org/. 

The SDSS is managed by the Astrophysical Research Consortium (ARC) for the
Participating Institutions. The Participating Institutions are The University of Chicago,
Fermilab, the Institute for Advanced Study, the Japan Participation Group, The Johns
Hopkins University, Los Alamos National Laboratory, the Max-Planck-Institute for
Astronomy (MPIA), the Max-Planck-Institute for Astrophysics (MPA), New Mexico State
University, University of Pittsburgh, Princeton University, the United States Naval
Observatory, and the University of Washington.

\appendix

\section{KL Gap-Correction in Quasar Spectra} \label{appendix:gapcorr}

\begin{figure}
\epsscale{0.8}
\plotone{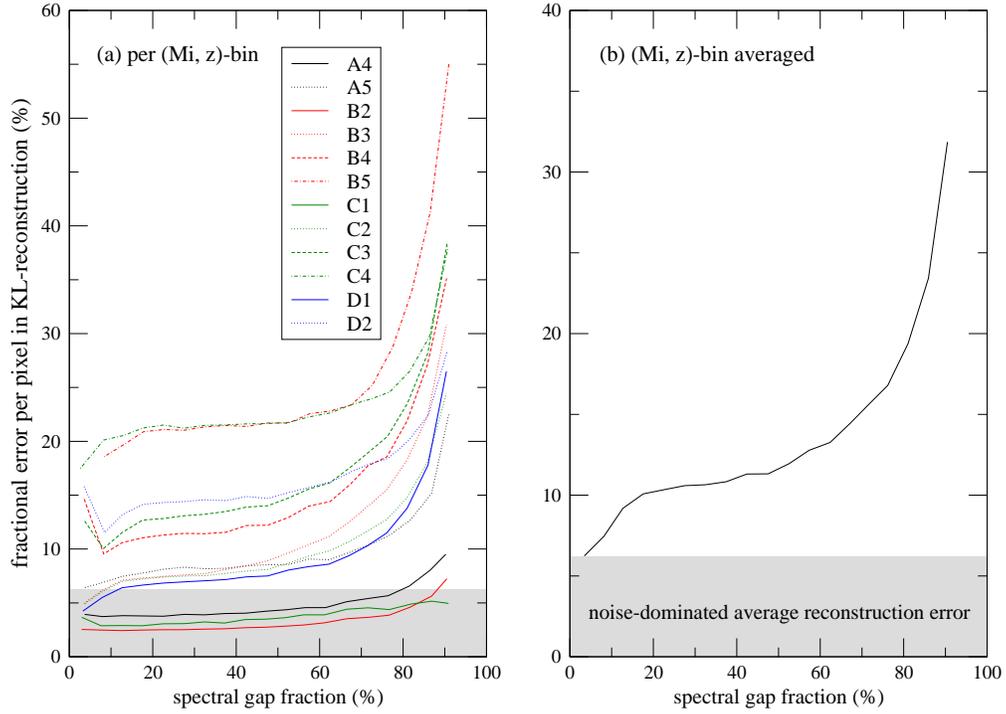}
\figcaption[fig26.eps]{The fractional error of KL-reconstruction
in flux density per wavelength bin as a function of the  
spectral gap fraction in the observed quasar spectra,  where in (a) each curve is averaged 
over all quasars in the corresponding $(M_{i}, z)$-bin, 
and in (b) the curve is averaged over all quasars in all $(M_{i}, z)$-bins. The gray region in each 
is the excluded region due to the average intrinsic noise per wavelength bin. \label{fig:ReconSpecErrVSGap}} 
\end{figure}

The construction of the $(M_{i}, z)$-bins in this work (\S~\ref{section:zbin}) is performed by constraining the gap 
fraction to be smaller than 50~\% for each spectrum to improve the accuracy of spectral reconstructions using 
eigenspectra. Here we discuss in detail how this value is arrived at. We artificially mask out (i.e. assign a 
zero weight) to given spectral intervals and study how well we can reconstruct these ``gappy'' regions from the 
eigenspectra \cite{Con99}. The comparison of the KL-reconstructed
spectrum with the original unmasked spectrum gives a direct assessment to the accuracy of the gap-correction 
procedure. We perform this test for the $(M_{i}, z)$-binned quasar spectra from this work. 
To simulate the effects of un-observed spectral regions due to different rest-wavelength 
coverage for quasars at different redshifts (the principal reason for gaps in the quasar spectra in our sample), 
each spectrum in all $(M_{i}, z)$-bins is artificially
masked at the short- and the long-wavelength ends. 
The masked spectra are then projected onto the
appropriate eigenspectra and the reconstructed spectra are
calculated using the first 50 modes. The fractional change in the flux density per wavelength bin 
(weighted by $w_{\lambda}$), $\sqrt{\sum_{\lambda} w_{\lambda} (f_{\lambda} - f_{\lambda}^{Recon})^{2}} / \sum_{\lambda} w_{\lambda} f_{\lambda}$, 
between the observed spectrum $f_{\lambda}$ and the reconstructed spectrum $f_{\lambda}^{Recon}$, 
averaged over all quasar spectra in each bin, 
are shown in Figure~\ref{fig:ReconSpecErrVSGap}a as a function of the spectral gap fraction.  
The gap fraction is calculated relative to the full restframe wavelength range, a variable for 
each quasar spectrum. The reconstruction from $50$ modes has an intrinsic error of approximately $6.2$~\% (due to 
the noise present in each spectrum, and the existence of $3.4$~\% bad pixels 
on average for each spectrum), which is estimated by reconstructing the spectra with no artificial gaps.
As expected, the difference between the unmasked observed spectrum and the reconstructed spectrum increases 
gradually with gap fraction.
 
Averaging over all $(M_{i}, z)$-bins (Figure~\ref{fig:ReconSpecErrVSGap}b), 
at a spectral gap fraction of $52.5$~\% 
the mean error in the 50-mode reconstruction is $\approx 11.9$~\%, which is $5.7$~\% above the 
noise-dominated average reconstruction error in the flux.
While a smaller gap fraction is in principle more desirable, 50~\%
is chosen to be the upper bound to compromise the fewer $(M_{i}, z)$-bins. 

In the construction of the global eigenspectra set covering the
rest-wavelength range $900-8000$~\AA, there are 89~\% of the QSOs
(Table~\ref{tab:gapglobal}) having spectral gap fractions larger than
50~\%. From Figure~\ref{fig:ReconSpecErrVSGap}b, we find that a gap
fraction larger than $76$~\% gives substantial reconstruction errors ($>
16.8$~\%), implying $\approx 17$~\% of the QSOs used in defining the
global eigenspectra may be poorly constrained when correcting for the
missing data. We stress that in defining the global eigenspectra from
the SDSS this is strictly the best estimation that can be made at
present, as no SDSS spectroscopic observations are available in the gap
regions at the red and the blue ends of the spectrum. The impact of this
gap correction is, as expected, wavelength dependent. Wavelengths
shortward of $5000$ \AA\ are very well constrained even with the global
eigenspectra with less than 1~\% of QSOs having gap corrections in excess
of 76~\% (Table~\ref{tab:gapglobal}). Determining the impact of the gaps
and the use of additional spectroscopic observations to complement the
SDSS data will be addressed in a future paper.

\begin{figure}
\epsscale{0.8}
\plotone{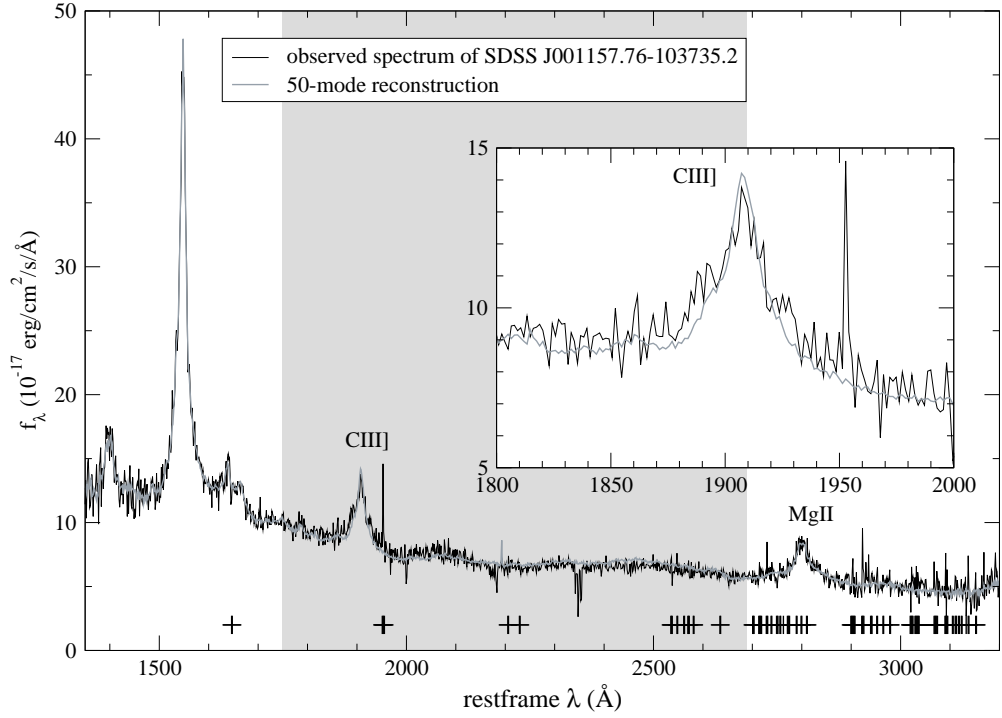}
\figcaption[fig27.eps]{A typical case of the 50-mode reconstruction using the $(M_{i}, z)$-binned 
eigenspectra. The spectrum has a total gap fraction of $56.5$~\% including the broad emission line
\ion{C}{3}$]$. The gray area is the artificially masked spectral region, and the 
crosses mark the bad pixels in the original spectrum.
The fractional reconstruction error per pixel is $12.1$~\%. The inset
shows the emission line \ion{C}{3}$]$ locally. \label{fig:recon_maskCIII_50percentGap}} 
\end{figure}

We also find that quasar broad emission lines can be reconstructed
locally using the $(M_{i}, z)$-binned eigenspectra with errors that are
typically small relative to the noise level.  For example, if
\ion{C}{3}$]$ is masked (over the region of influence $1830-1976$~\AA),
averaging over all QSOs in the bins {\bf B3} and {\bf C3}, the 50-mode
reconstruction error described above is 10.4~\%; and for \ion{Mg}{2}
(over the region of influence $2686-2913$~\AA), 11.3~\%.  For the case in
which at least one broad emission line is masked  and with a substantial total gap fraction (in our case, \ion{C}{3}$]$; and a mean spectral gap fraction of $60.0$~\%), the average
reconstruction error per pixel is found to be $12.5$~\%  when averaging over the bins {\bf B3}
and {\bf C3}.  Figure~\ref{fig:recon_maskCIII_50percentGap} shows the
observed and the reconstructed spectra of an object with a
reconstruction error approximately equal to the average 
value. While the reconstructed continuum has a small difference from the
observed continuum, the emission line \ion{C}{3}$]$ is reconstructed
well, extremely well if considering the fact that the whole region of influence is
within the masked region. The actual quality of the reconstruction depends on the individual
spectrum and position and size of the gaps.

\clearpage

%______________________________________________________________________
%______________________________________________________________________

%______________________________________________________________________
%______________________________________________________________________

\clearpage

\clearpage

\clearpage

\clearpage

\clearpage

\clearpage

\clearpage

\clearpage

\clearpage

\clearpage

\clearpage

\clearpage

\clearpage

\clearpage

\clearpage

\clearpage

\clearpage

\clearpage

\clearpage

\clearpage

\clearpage

\clearpage

\clearpage

\clearpage

\clearpage

\clearpage

\clearpage

\clearpage
\begin{table}
\caption{The partial sums of weights of the global QSO eigenspectra. 
\label{tab:weights_all}} 
\begin{center}
\begin{tabular}{cc}
\hline
\hline  
 Number of first m-modes: m & Weight \\
\hline
  1 &   0.560887 \\
  2 &   0.680197 \\
  3 &   0.755992 \\
  4 &   0.822280 \\ 
  5 &   0.849927 \\
  8 &   0.896213 \\
 10 &   0.919394 \\
 15 &   0.953041 \\
 20 &   0.968920 \\
 50 &   0.995680 \\
 75 &   0.998512 \\
100 &   0.999199 \\
\hline
\end{tabular}
\end{center}
\end{table}

\clearpage

\begin{table}
\caption{The subsamples for performing the commonality analysis on the resultant sets of eigenspectra.}
\label{tab:subsample}
\begin{center}
\begin{tabular}{cccc}
\hline
\hline
 Name & Redshift range & Luminosity range ($M_{i}$) & Number of objects  \\
\hline  
 Subsample 1 & 0.9 to 1.1 & -24 to -25 & 472 \\
 Subsample 2 & 0.9 to 1.1 & -24 to -25 & 236 \\
 Subsample 3 & 0.9 to 1.1 & -25 to -26 & 442 \\
 Subsample 4 & 1.1 to 1.3 & -25 to -26 & 469 \\ 
\hline
\end{tabular}
\end{center}
\end{table}

\clearpage

\begin{table}
\caption{The number of QSOs in the $(M_{i}, z)$-bins. 
	\label{tab:cuts}}
\begin{center}
{\small
\begin{tabular}{rcccc}
\hline
                        & {\bf A} & {\bf B} & {\bf C} & {\bf D} \\
 & $M_{i} = (-30, -28)$ &  $M_{i} = (-28, -26)$  &  $M_{i} = (-26, -24)$  &  
$M_{i} = (-24, -22)$  \\
\hline
\hline
{\bf ZBIN~1}: $0.08< z <0.53$  &      &      &                      109 & 1597
 \\
              $2486-8000$~\AA  &      &      &  ({{95\%}\tablenotemark{a}~, {0\%}\tablenotemark{b}}~) & (81\%, 4\%)
 \\
\hline
   {\bf  2}:  $0.53< z <1.16$  &      &          178 &         2752 & 1351
 \\
              $1759-6018$~\AA  &      &  (94\%, 0\%) & (79\%, 12\%) & (30\%, 45\%)
 \\
\hline
   {\bf  3}:  $1.16< z <2.06$  &      &         3477 &         4462 & 
 \\
              $1242-3800$~\AA  &      &  (92\%, 4\%) & (41\%, 46\%) & 
 \\
\hline
   {\bf  4}:  $2.06< z <3.33$  &          110 &         1796 &         477 & 
 \\
               $900-3005$~\AA  &  (74\%, 0\%) & (65\%, 11\%) & (0\%, 61\%) & 
 \\
\hline
   {\bf  5}:  $3.33< z <5.13$  &            85 &          352 &      &  
 \\
               $900-2123$~\AA  &   (94\%, 0\%) &  (75\%, 1\%) &      & 
 \\
\hline  
\end{tabular}
}
\end{center}
\tablenotetext{a}{The percentage of quasars (to the nearest unity) that are targeted by the quasar multi-dimensional
color-space in the SDSS \cite{Richards02a}.}
\tablenotetext{b}{The percentage of quasars targeted solely by the SDSS Serendipity module.}
\end{table}

\clearpage

\begin{table}
\caption{The partial sums of weights of the $(M_{i},z)$-binned QSO eigenspectra.
	\label{tab:weights}}
\begin{center}
{\small
\begin{tabular}{rcccccccc}
\hline
weight & 1 mode & 2 modes & 3 modes  & 5 modes & 10 modes & 15 modes & 20 modes & 50 modes \\
\hline  
\hline  
{\bf ZBIN}: {\bf 1} & 0.9284 & 0.9646 & 0.9729 & 0.9836 & 0.9915 & 0.9934 & 0.9942 & 0.9945 \\ 
            {\bf 2} & 0.9317 & 0.9657 & 0.9752 & 0.9815 & 0.9874 & 0.9896 & 0.9905 & 0.9909 \\
            {\bf 3} & 0.9232 & 0.9556 & 0.9651 & 0.9763 & 0.9841 & 0.9870 & 0.9880 & 0.9885 \\
            {\bf 4} & 0.8737 & 0.9089 & 0.9298 & 0.9455 & 0.9608 & 0.9685 & 0.9719 & 0.9738 \\
            {\bf 5} & 0.8122 & 0.8540 & 0.8783 & 0.8986 & 0.9247 & 0.9356 & 0.9398 & 0.9422 \\
\hline
\hline  
 $(M_{i}, z)$-bin: {\bf A4} & 0.9134 & 0.9449 & 0.9551 & 0.9663 & 0.9781 & 0.9842 & 0.9866 & 0.9881 \\
                   {\bf A5} & 0.8801 & 0.9076 & 0.9242 & 0.9384 & 0.9555 & 0.9637 & 0.9676 & 0.9699 \\
\hline
{\bf B2} & 0.9789 & 0.9926 & 0.9958 & 0.9973 & 0.9984 & 0.9989 & 0.9991 & 0.9991 \\
{\bf B3} & 0.9474 & 0.9701 & 0.9781 & 0.9859 & 0.9911 & 0.9931 & 0.9938 & 0.9941 \\
{\bf B4} & 0.8794 & 0.9115 & 0.9292 & 0.9516 & 0.9673 & 0.9743 & 0.9771 & 0.9787 \\
{\bf B5} & 0.8083 & 0.8487 & 0.8718 & 0.8941 & 0.9207 & 0.9314 & 0.9358 & 0.9383 \\
\hline
{\bf C1} & 0.9893 & 0.9955 & 0.9969 & 0.9981 & 0.9991 & 0.9994 & 0.9995 & 0.9995 \\
{\bf C2} & 0.9470 & 0.9709 & 0.9819 & 0.9870 & 0.9921 & 0.9938 & 0.9944 & 0.9947 \\
{\bf C3} & 0.9226 & 0.9546 & 0.9631 & 0.9739 & 0.9805 & 0.9832 & 0.9842 & 0.9848 \\
{\bf C4} & 0.8365 & 0.8756 & 0.8983 & 0.9191 & 0.9387 & 0.9494 & 0.9539 & 0.9565 \\
\hline
{\bf D1} & 0.9289 & 0.9640 & 0.9715 & 0.9829 & 0.9908 & 0.9928 & 0.9937 & 0.9941 \\
{\bf D2} & 0.9086 & 0.9469 & 0.9586 & 0.9698 & 0.9768 & 0.9797 & 0.9809 & 0.9816 \\
\hline
\end{tabular}
}
\end{center}
\end{table}

\clearpage

\begin{table}
\caption{The average FWHMs of major broad emission lines of the {\bf ZBIN}s QSO eigenspectra.
\label{tab:FWHMzbin}}
\begin{tabular}{lcccccc}
\hline
\hline
                                     &   & &     & FWHM\tablenotemark{a} (km s$^{-1}$)    &         &  \\
                                     &   & &     &              ({\bf ZBIN}) &         &  \\
Percentage of Lines\tablenotemark{b} & 1st$-$3rd & 1st$-$2nd & 1st & 1st+2nd\tablenotemark{c} & 1st+3rd & \\
\hline
Narrower than 1st mode & 77\% & 40\% & \nodata & 61\% & 39\% & \\
   Wider than 1st mode & 23\% & 60\% & \nodata & 39\% & 61\% & \\
\hline 
Example: & & & & & & Redshift bin \\
\hline 
Ly$\alpha$+N~V $(1160, 1290)$\AA\tablenotemark{d} &
8999 & 3343 & 4405 & 7918\tablenotemark{e} & 3116
& {\bf ZBIN~5} \\
C~IV $(1494, 1620)$\AA &
3579 & 4110 & 4140 & 4179 & 4913
& {\bf ZBIN~4}  \\
C~III$]$ $(1830, 1976)$\AA &
5291 & 5597 & 5905 & 6296 & 6588     
& {\bf ZBIN~4}  \\
Mg~II $(2686, 2913)$\AA & 
4274 & 4318 & 4204 & 4096 & 4126   
& {\bf ZBIN~3}  \\
H$\beta$ $(4050, 4152)$\AA &
1661 & 2035 & 1795 & 1443 & 1998   
& {\bf ZBIN~1}  \\
$[$O~III$]$$\lambda5008$ $(4982, 5035)$\AA &  
493 & 494 & 495 & 495 & 497    
& {\bf ZBIN~1}  \\
\hline
\end{tabular}
\tablenotetext{a}{The values are to the nearest unity.}
\tablenotetext{b}{Only the emission lines with line-widths $> 1000$~km~s$^{-1}$ are counted, 
in both cases of reconstructions using the first mode and the concerned linear-combination.}
\tablenotetext{c}{The expansion coefficients in the linear-combination are taken to be the (signed) medians of
the eigencoefficients of all objects in the concerned sample.} 
\tablenotetext{d}{The restframe wavelength window across which the continuum underneath the line is approximated by 
linear-interpolation.} 
\tablenotetext{e}{Though not completely deblended from N~V, the 2nd eigenspectrum mainly contains the 
Ly$\alpha$ component.} 
\end{table}

\clearpage

\begin{deluxetable}{rlcccclll}
\tabletypesize{\tiny}
\rotate
\tablewidth{0pt}
\tablecaption{Correlations among major emission lines deduced from the local QSO eigenspectra sets.}
\tablehead{
\colhead{Line} &  \colhead{$\lambda_{lab}$ (\AA)} & \colhead{$(\lambda_{low}, \lambda_{upp})$\tablenotemark{a}} (\AA) & 
\colhead{$\lambda_{rest}$\tablenotemark{b} (\AA)} & \colhead{$z$ (num. of obj.)\tablenotemark{b}} & 
\colhead{EW$_{rest}$\tablenotemark{c} (\AA)} & \colhead{Reg. Coeff.\tablenotemark{f}} & \colhead{Corr. Coeff.\tablenotemark{f}} & \colhead{P-value\tablenotemark{g}}}
\startdata
          Ly$\beta$+OVI & 1025.72+1033.83 & $(1012, 1055)$ & $900-1320$ & $3.22-5.414~(496)$ & $9.08 - 14.30$  & $1.0000$  & $1.0000$  & 0.000000 \\
          Ly$\alpha$+NV & 1215.67+1240.14 & $(1160, 1290)$ & & & $74.17 - 140.28$  & $12.6411$  & $0.9985$  & 0.000000 \\
                             NV & 1240.14 & $(1230, 1252)$ & & & $2.31 - 1.05$  & $-0.2405$  & $0.9332$  & 0.000080 \\
                OI+SiII & 1304.35+1306.82 & $(1290, 1318)$ & & & $1.10 - 2.59$  & $0.2833$  & $0.9931$  & 0.000000 \\
\hline
          Ly$\alpha$+NV & 1215.67+1240.14 & $(1160, 1290)$ & $1150-2000$ & $2.3-3.6~(1277)$  & $122.96 - 65.82$  & $1.0000$  & $1.0000$  & 0.000000 \\
                             NV & 1240.14 & $(1230, 1252)$ & & & $1.87 - 2.06$  & $-0.0032$  & $0.9760$  & 0.000001 \\
                OI+SiII & 1304.35+1306.82 & $(1290, 1318)$ & & & $2.40 - 1.22$  & $0.0206$  & $0.9999$  & 0.000000 \\
                            CII & 1335.30 & $(1325, 1348)$ & & & $0.72 - 0.67$  & $0.0009$  & $0.9862$  & 0.000000 \\
            SiIV+OIV$]$ & 1396.76+1402.06 & $(1360, 1446)$ & & & $9.12 - 9.24$  & $-0.0021$  & $0.9812$  & 0.000001 \\
                            CIV & 1549.06 & $(1494, 1620)$ & & & $35.45 - 17.56$  & $0.3129$  & $0.9997$  & 0.000000 \\
                           HeII & 1640.42 & $(1622, 1648)$ & & & $1.06 - 0.35$  & $0.0125$  & $0.9921$  & 0.000000 \\
OIII$]$+AlII+FeII(UV40) & 1664.74\tablenotemark{d} & $(1648, 1682)$ & & & $1.24 - 0.13$  & $0.0195$  & $0.9534$  & 0.000020 \\
                        NIII$]$ & 1750.26 & $(1735, 1765)$ & & & $0.74 - 0.30$  & $0.0077$  & $0.9966$  & 0.000000 \\
   FeII(UV191) & 1788.73\tablenotemark{d} & $(1771, 1802)$ & & & $0.59 - 0.25$  & $0.0059$  & $0.9978$  & 0.000000 \\
                          AlIII & 1857.40 & $(1840, 1875)$ & & & $0.49 - 0.66$  & $-0.0030$  & $0.9601$  & 0.000011 \\
               CIII$]$+Iron lines & 1905.97\tablenotemark{d} & $(1830, 1976)$ & & & $24.93 - 21.82$  & $0.0541$  & $0.9891$  & 0.000000 \\
\hline
                          AlIII & 1857.40 & $(1840, 1875)$ & $1800-3200$ & $1.1-1.87~(6998)$  & $0.44 - 0.53$  & $1.0000$  & $1.0000$  & 0.000000 \\
               CIII$]$+Iron lines & 1905.97\tablenotemark{d} & $(1830, 1976)$ & & & $25.47 - 23.14$  & $-26.7053$  & $0.9962$  & 0.000000 \\
                    FeIII(UV48) & 2076.62 & $(2036, 2124)$ & & & $2.67 - 3.06$  & $4.4128$  & $0.9999$  & 0.000000 \\
                         CII$]$ & 2326.44 & $(2312, 2338)$ & & & $0.42 - 0.35$  & $-0.8192$  & $0.9933$  & 0.000000 \\
$[$NeIV$]$+FeIII(UV47) & 2423.46\tablenotemark{d} & $(2402, 2448)$ & & & $0.83 - 0.85$  & $0.1746$  & $0.9987$  & 0.000000 \\
                           MgII & 2798.75 & $(2686, 2913)$ & & & $35.81 - 33.98$  & $-20.8265$  & $0.9973$  & 0.000000 \\
  OIII+FeII(Opt82) & 3127.70\tablenotemark{d} & $(3100, 3153)$ & & & $0.66 - 0.48$  & $-1.9915$  & $0.9883$  & 0.000000 \\
\hline
                           MgII & 2798.75 & $(2686, 2913)$ & $2600-4250$ & $0.46-1.16~(4647)$ & $34.97 - 42.33$  & $1.0000$  & $1.0000$  & 0.000000 \\
  OIII+FeII(Opt82) & 3127.70\tablenotemark{d} & $(3100, 3153)$ & & & $1.34 - 1.05$  & $-0.0403$  & $0.9903$  & 0.000000 \\
                      $[$NeV$]$ & 3346.82 & $(3329, 3356)$ & & & $0.60 - 0.22$  & $-0.0522$  & $0.9404$  & 0.000051 \\
                      $[$NeV$]$ & 3426.84 & $(3394, 3446)$ & & & $2.04 - 0.99$  & $-0.1434$  & $0.9619$  & 0.000009 \\
                      $[$OII$]$ & 3728.48 & $(3714, 3740)$ & & & $3.36 - 0.71$  & $-0.3591$  & $0.9041$  & 0.000329 \\
                    $[$NeIII$]$ & 3869.85 & $(3850, 3884)$ & & & $2.34 - 0.87$  & $-0.1986$  & $0.9441$  & 0.000040 \\
$[$NeIII$]$+H$\epsilon$ & 3968.58+3971.20 & $(3950, 3978)$ & & & $0.73 - 0.33$  & $-0.0533$  & $0.9593$  & 0.000011 \\
                      H$\delta$ & 4102.89 & $(4050, 4152)$ & & & $4.83 - 6.32$  & $0.2001$  & $0.9996$  & 0.000000 \\
\hline
                      H$\delta$ & 4102.89 & $(4050, 4152)$ & $4000-5500$ & $0.08-0.67~(2534)$ & $6.96 - 5.30$  & $1.0000$  & $1.0000$  & 0.000000 \\
                      H$\gamma$ & 4341.68 & $(4285, 4412)$ & & & $14.72 - 11.08$  & $2.1956$  & $1.0000$  & 0.000000 \\
                     $[$OIII$]$ & 4364.44 & $(4352, 4372)$ & & & $1.09 - 0.24$  & $0.5133$  & $0.9520$  & 0.000022 \\
   FeII(optical, blended lines)\tablenotemark{e} & 4600.00 & $(4469, 4762)$ & & & $14.69 - 21.28$  & $-3.9796$  & $0.9795$  & 0.000001 \\
                           HeII & 4687.02 & $(4668, 4696)$ & & & $0.99 - 0.03$  & $0.5797$  & $0.8904$  & 0.000551 \\
                       H$\beta$ & 4862.68 & $(4760, 4980)$ & & & $59.38 - 43.26$  & $9.7307$  & $0.9999$  & 0.000000 \\
                     $[$OIII$]$ & 4960.30 & $(4945, 4972)$ & & & $10.82 - 0.08$  & $6.4826$  & $0.8847$  & 0.000671 \\
                     $[$OIII$]$ & 5008.24 & $(4982, 5035)$ & & & $40.46 - 2.64$  & $22.8252$  & $0.9063$  & 0.000301 \\
   FeII(optical, blended lines)\tablenotemark{e} & 5250.00 & $(5100, 5477)$ & & & $13.80 - 26.49$  & $-7.6594$  & $0.9600$  & 0.000011 \\
$[$FeVII$]$+FeII(Opt49) & 5277.92\tablenotemark{d} & $(5273, 5287)$ & & & $0.16 - 0.13$  & $0.0204$  & $0.9999$  & 0.000000 \\
\hline
$[$FeVII$]$+FeII(Opt49) & 5277.92\tablenotemark{d} & $(5273, 5287)$ & $5200-7000$ & $0.08-0.31~(456)$  & $0.15 - 0.18$  & $1.0000$  & $1.0000$  & 0.000000 \\
                            HeI & 5877.29 & $(5805, 5956)$ & & & $8.60 - 3.09$  & $-258.9123$  & $0.9423$  & 0.000045 \\
    H$\alpha$+$[$NII$]$ & 6564.61+6585.28 & $(6400, 6765)$ & & & $340.52 - 138.68$  & $-9493.8027$  & $0.9525$  & 0.000021 \\
                      $[$NII$]$ & 6585.28 & $(6577, 6593)$ & & & $1.21 - 2.63$  & $65.5191$  & $0.9796$  & 0.000001 \\
                      $[$SII$]$ & 6718.29 & $(6708, 6726)$ & & & $1.28 - 1.13$  & $-7.2297$  & $0.9967$  & 0.000000 \\
                      $[$SII$]$ & 6732.67 & $(6726, 6742)$ & & & $0.91 - 0.65$  & $-12.5154$  & $0.9888$  & \\
\hline
                      $[$NII$]$ & 6585.28 & $(6577, 6593)$ & $6500-8000$ & $0.08-0.15~(29)$   & $2.96 - 2.39$  & $1.0000$  & $1.0000$  & 0.000000 \\
                      $[$SII$]$ & 6718.29 & $(6708, 6726)$ & & & $1.30 - 1.10$  & $0.3411$  & $0.9998$  & 0.000000 \\
                      $[$SII$]$ & 6732.67 & $(6726, 6742)$ & & & $0.62 - 0.94$  & $-0.5424$  & $0.9800$  & 0.000001 \\
                    $[$ArIII$]$ & 7137.80 & $(7131, 7148)$ & & & $0.63 - 0.11$  & $0.8965$  & $0.9424$  & 0.000045 \\
\enddata
\tablenotetext{a}{$(\lambda_{low}, \lambda_{upp})$ is the rest-wavelength window within which the local continuum 
is estimated by linear-interpolation. These wavelength windows are determined by Vanden Berk {et~al.} (2001).}
\tablenotetext{b}{The restricted wavelengths, the corresponding redshifts and the numbers of the quasars 
chosen for the KL transforms. In a given redshift range, 
all appropriate quasars in our sample are included regardless of their absolute magnitudes,
hence the correlations listed here are the ensemble-averaged properties.}
\tablenotetext{c}{The range of restframe equivalent widths of the emission line along the spectral
sequence $(a_1, a_2)$ (in decreasing $a_2$ values).}
\tablenotetext{d}{The observed wavelengths ($\lambda_{obs}$; in vacuum) as listed in Vanden Berk {et~al.} (2001).}
\tablenotetext{e}{Uncertain $\lambda_{lab}$ and $(\lambda_{low}, \lambda_{upp})$ due to the unknown number 
of lines and their blended nature.} 
\tablenotetext{f}{The linear regression and the linear correlation coefficients are given in 4 significant figures.} 
\tablenotetext{g}{The two-tailed P-value for the correlation coefficient.} \label{tab:linedata}
\end{deluxetable}

\clearpage

\begin{table}
\caption{The average FWHMs of major broad emission 
lines of the local QSO eigenspectra.
\label{tab:FWHMlocal}}
\begin{tabular}{lcccccc}
\hline
\hline
                                     & & &     & FWHM (km s$^{-1}$) &         \\
                                     & & &     &            (local) &         \\
     Percentage of Lines             & 1st$-$3rd & 1st$-$2nd & 1st &      1st+2nd\tablenotemark{a} & 1st+3rd & \\
\hline
Narrower than the 1st mode & 72\% & 31\% & \nodata & 76\% & 52\% & \\
   Wider than the 1st mode & 28\% & 69\% & \nodata & 24\% & 48\% & \\
\hline
Example: & & & & & & redshifts\\
\hline
Ly$\alpha$+N~V $(1160, 1290)$\AA &  
2789 & 9139 & 3820 & 3104\tablenotemark{b} & 5549 & $2.3-3.6$ \\
C~IV $(1494, 1620)$\AA &  
3000 & 4594 & 3763 & 3339 & 4778 & $2.3-3.6$ \\
C~III$]$ $(1830, 1976)$\AA &  
5904 & 6009 & 5802 & 5552 & 5721 & $1.1-1.87$ \\
Mg~II $(2686, 2913)$\AA &  
4194 & 4171 & 4149 & 4113 & 4117 & $1.1-1.87$ \\
H$\beta$ $(4050, 4152)$\AA &  
1978 & 2238 & 1997 & 1584 & 1999 & $0.08-0.67$ \\
$[$O~III$]$$\lambda$5008 $(4982, 5035)$\AA &  
551 & 571 & 535 & 521 & 522 & $0.08-0.67$ \\
\hline
\end{tabular}
\tablenotetext{a}{The 2nd eigenspectrum shows low-velocity (line-core) components of the broad emission lines.}
\tablenotetext{b}{The Ly$\alpha$ and N~V lines are not deblended in the 1st order (i.e., the mean spectrum),
and are in the 2nd order, meaning that it is partly due to the  deblending which causes the
narrower width in the 2nd mode.}
\end{table}

\begin{table}
\caption{The average FWHMs of major broad emission lines of the global QSO eigenspectra.}
\label{tab:FWHMglobal}
\begin{tabular}{lccccc}
\hline
\hline
                                     &   & &     & FWHM (km s$^{-1}$)    &         \\
                                     &   & &     &              (global) &          \\
      Percentage of Lines            & 1st$-$3rd & 1st$-$2nd & 1st & 1st+2nd & 1st+3rd \\
\hline
Narrower than 1st mode & 26\% & 50\% & \nodata & 48\% & 73\% \\
   Wider than 1st mode & 74\% & 50\% & \nodata & 52\% & 27\% \\
\hline
Example: & & & & & \\
\hline
Ly$\alpha$+N~V $(1160, 1290)$\AA & 
 6837 & 5045 & 5103 & 9105\tablenotemark{a} & 3678 \\
C~IV $(1494, 1620)$\AA & 
 4787 & 4536 & 4518 & 4485 & 4341  \\
C~III$]$ $(1830, 1976)$\AA & 
 6405 & 5867 & 6004 & 6195 & 5626  \\
Mg~II $(2686, 2913)$\AA & 
 4035 & 3997 & 3980 & 3959 & 3918  \\
H$\beta$ $(4050, 4152)$\AA & 
 2605 & 3090 & 2506 & 2244 & 2351  \\
$[$O~III$]$$\lambda$5008 $(4982, 5035)$\AA &  
! 566 & 525 & 546 & 551 & 538  \\
\hline
\end{tabular}
\tablenotetext{a}{The Ly$\alpha$ and N~V are deblended in the 2nd eigenspectrum and are blended in the 1st one.}
\end{table}

\clearpage

\begin{deluxetable}{cccc}
\tablewidth{0pt} 
\tablecaption{Spectral gap fraction of the whole sample.}
\tablehead{\colhead{Spectral gap fraction larger than} & \multicolumn{3}{c}{Number (fraction) of QSOs} \\ \cline{2-4} \\
\colhead{$\lambda_{rest}$-range of the eigenspectra:} & \colhead{$900-8000$~\AA} & \colhead{$900-7000$~\AA} & 
\colhead{$900-5000$~\AA} }
\startdata 
 0.4 & 16,420 (0.98) & 15,313 (0.92) & 10,423 (0.62) \\
 0.5 & 15,050 (0.90) & 13,561 (0.81) &  6,421 (0.38) \\
 0.6 & 12,696 (0.76) & 10,275 (0.62) &  1,682 (0.10) \\
 0.7 &  7,424 (0.44) &  3,519 (0.21) &    423 (0.025) \\
0.75 &  2,920 (0.17) & 1,131 (0.068) &    100 (0.0060) \\
 0.8 &   873 (0.052) &   416 (0.025) &      0 (0.00) \\
 0.9 &      0 (0.00) &      0 (0.00) &      0 (0.00) \\
\enddata 
\label{tab:gapglobal}
\end{deluxetable}

\end{document}